\documentstyle[12pt,axodraw]{article}
\def\lesssim{\mathrel{\mathpalette\vereq<}}
\def\gtrsim{\mathrel{\mathpalette\vereq>}}

\def\Im{\mbox{Im}\,}
\makeatletter
\def\vereq#1#2{\lower3pt\vbox{\baselineskip1.5pt \lineskip1.5pt
\ialign{$\m@th#1\hfill##\hfil$\crcr#2\crcr\sim\crcr}}}
\makeatother

\begin{document}

\begin{titlepage}

\begin{flushright}
SNS-PH/00-16 \\
UCB-PTH-00/38 \\
LBNL-47128 \\
\end{flushright}

\vskip 1.5cm

\begin{center}
{\Large \bf  A Constrained Standard Model \\
             from a Compact Extra Dimension}

\vskip 1.0cm

{\bf
Riccardo Barbieri$^{a}$,
Lawrence J.~Hall$^{b,c}$,
Yasunori Nomura$^{b,c}$
}

\vskip 0.5cm

$^a$ {\it Scuola Normale Superiore and INFN, Piazza dei Cavalieri 7, 
                 I-56126 Pisa, Italy}\\
$^b$ {\it Department of Physics, University of California,
                 Berkeley, CA 94720, USA}\\
$^c$ {\it Theoretical Physics Group, Lawrence Berkeley National Laboratory,
                 Berkeley, CA 94720, USA}

\vskip 1.0cm

\abstract{A $SU(3) \times SU(2) \times U(1)$ supersymmetric theory is 
constructed with a TeV sized extra dimension compactified on the orbifold 
$S^1/(Z_2 \times Z_2')$.  The compactification breaks supersymmetry 
leaving a set of zero modes which correspond precisely to the states of 
the 1 Higgs doublet standard model.  Supersymmetric Yukawa interactions 
are localized at orbifold fixed points.  The top quark hypermultiplet 
radiatively triggers electroweak symmetry breaking, yielding a Higgs 
potential which is finite and exponentially insensitive to physics above 
the compactification scale.  This potential depends on only a single 
free parameter, the compactification scale, yielding a Higgs mass 
prediction of $127 \pm 8$ GeV.  The masses of the all superpartners, and 
the Kaluza-Klein excitations are also predicted.  The lightest 
supersymmetric particle is a top squark of mass $197 \pm 20$ GeV. 
The top Kaluza-Klein tower leads to the $\rho$ parameter having quadratic 
sensitivity to unknown physics in the ultraviolet.}

\end{center}
\end{titlepage}

\section{Introduction}  \label{sec:intro}

The standard model provides an economical description of particles and
their interactions in terms of 18 free parameters. There are 9
parameters associated with the masses of the quarks and charged
leptons, and 4 to describe the flavor mixing of the quarks. There are
3 independent gauge couplings, and the final 2 parameters are 
associated with the Higgs boson. One is the vacuum expectation value 
(VEV) of the Higgs field, which is accurately determined
by the Fermi coupling constant, and the other is the mass of the Higgs
boson, which is unknown, although precision electroweak data suggests 
it is less than 188 GeV at 95\% confidence level \cite{Higgs}.

Despite the phenomenological success and mathematical consistency of
the standard model, it does not provide a physical description of
electroweak symmetry breaking (EWSB). The theory is believed to be an
effective theory valid at all energies below some cutoff,
$\Lambda$. Yet the mass parameter of the Higgs field has radiative
corrections that grow quadratically with $\Lambda$
\begin{equation}
m_H^2 \propto - \Lambda^2.
\label{eq:quaddiv}
\end{equation}
The physics of EWSB is at or beyond the cutoff, and hence not 
adequately described by the low energy effective theory.

In this paper we introduce a theory that does provide a full physical
description of EWSB, in terms of new physics at a mass scale
of 400 GeV. Our theory contains 17 free parameters, one fewer than the
standard model, so that we are able to predict the mass of the Higgs
boson. 

A key feature of most theories which go beyond the standard model is
an enhanced symmetry structure. Symmetries are the key to constructing
more predictive and elegant theories. Yet they are also a challenge,
because nature does not possess these additional symmetries, so they
must be broken and this often introduces great freedom. 
For example, grand unified theories provide a relation
between the three gauge coupling constants, and yet breaking the grand
unified gauge symmetry is the least attractive and constrained aspect
of these theories.

It is commonly believed that the most satisfactory way to construct a
physical theory of EWSB is to extend spacetime symmetry to include
supersymmetry \cite{SUSY}. In this case, the quadratic divergence in 
the Higgs boson mass parameter coming from a top quark radiative 
correction is cancelled by that coming from a scalar top. 
Including supersymmetry breaking at scale $m_{\rm SUSY}$, the resulting 
divergence is logarithmic
\begin{equation}
m_H^2 \propto - m_{\rm SUSY}^2 \ln \Lambda,
\label{eq:logdiv}
\end{equation}
so that EWSB may be triggered by physics all the way up to the 
cutoff \cite{radewsb}. The Higgs mass is reliably computed 
in the effective theory, and is not dominated by unknown
physics at the cutoff. However, the  economy of the theory has been
sacrificed. An entirely new sector of the theory must be introduced,
with the sole purpose of breaking supersymmetry to generate the scale 
$m_{\rm SUSY}$. 
There are now many alternatives for this sector, and their
relative merits are hotly debated, but the fact remains that such a
new sector of the theory is inherent to the present formulation of
supersymmetric theories. Indeed it was realized from an early stage
that supersymmetry breaking could not occur in the standard model
sector, but had to be somewhat remote from it \cite{DG}, necessitating a
mediation mechanism between the two sectors.

The result of this mediation is to introduce a set of new parameters
describing the strengths of the soft supersymmetry breaking
interactions. These parameters themselves give rise to a host of new
problems: why are the squarks nearly degenerate so as to avoid
flavor-changing and CP violating problems? Why are there any light
Higgs bosons in the theory? There is no obvious symmetry keeping them
light. Why is the proton stable? This success of the standard model is
lost when the theory is made supersymmetric --- apparently the proton
could decay via squark exchange. What distinguishes matter from Higgs?
In the standard model this is clear: matter is fermionic while Higgs
is bosonic. Supersymmetry provides no such clear separation,
requiring an artificial distinction between the Higgs
boson and the sneutrino. Studying solutions to these problems has 
been an active area of research for many years.

Finally, one can ask how well supersymmetric theories account for the
mass scale of the weak interactions, given that no superpartners have
been discovered. The naturality of the proposed models is certainly
not perfect: in gravity mediation only about 3\% of parameter space
gives acceptable theories, while in other schemes, such as gauge
mediation, a similar amount of tuning is required to keep the charged 
slepton masses above the LEP2 limit.

On the other hand,
supersymmetric theories with a perturbative energy desert certainly 
have some very positive features. It is
possible to construct a relatively complete framework with a
successful, precise prediction of the weak mixing angle \cite{DG, S-SG}, 
the correct order of magnitude for neutrino masses from the see-saw 
mechanism \cite{Seesaw}, and the correct order of magnitude 
for the dark matter abundance from the cosmological freezeout of 
the lightest supersymmetric particle (LSP) \cite{DM}.
However, given the shortcomings discussed above, we are motivated to
investigate a more economical theory of EWSB. 

Consider compact spatial dimensions, with a compactification scale
$R^{-1}$ of order an inverse TeV, 
in which standard model particles propagate \cite{antoniadis}. 
In this case there are Kaluza-Klein (KK) towers for each particle
propagating in this bulk. 
Imposing a symmetry on a compact space reduces the number of modes in 
the KK tower. This orbifold construction is crucial in obtaining chiral
zero mode quarks, from a 5d theory which is vector-like \cite{orbifold}.
Furthermore, assume that the underlying bulk theory is supersymmetric. 
For example, if the top quark propagates in the bulk there will be 
KK towers for both the top quark and the top squark.
In such a situation one can study radiative corrections to the Higgs
boson mass with contributions from the entire KK towers \cite{KKEWSB}. 
In the case that supersymmetry is unbroken, there is
an exact cancellation between the top and stop KK towers. 
However, if supersymmetry is broken so that the masses of the squark
tower are shifted relative to the masses of the quark tower by an
amount of order $R^{-1}$, the cancellation is no longer
complete. Remarkably, the result is completely finite:
\begin{equation}
m_H^2 \propto - \left( {1 \over R} \right)^2,
\label{eq:finite}
\end{equation}
and therefore independent of the cutoff, $\Lambda$, of the theory.
The introduction of an extra compact dimension at the TeV scale 
allows a new resolution of the Higgs mass divergence ---
one where the physics of EWSB is necessarily right at the weak scale
itself \cite{KKEWSB}. Such Kaluza-Klein EWSB implies that the gauge couplings
will become non-perturbative not far above the TeV scale \cite{DDG} 
--- there is no perturbative energy desert --- and fits well with the 
possibility that the fundamental scale of gravity is in the multi-TeV
domain \cite{ADD}. What breaks supersymmetry, causing the mass shift between
quark and squark KK towers? 

Once extra dimensions have been introduced at the TeV scale, 
a new possibility opens up for supersymmetry breaking 
\cite{antoniadis, SS-SB}, that is not available in the 
conventional energy desert version of supersymmetry: 
the Scherk-Schwarz mechanism \cite{SS}. 
Modifying the periodic boundary condition by using an $R$ symmetry, 
the excluded modes are different for fermions and bosons, 
breaking supersymmetry. 
An explicit extension of the standard model which breaks
supersymmetry via the Scherk-Schwarz mechanism has been
proposed \cite{DPQ}, illustrating some important advantages over
conventional supersymmetric theories. For example, the $\mu$ problem
is solved, and supersymmetry breaking generates tree level Dirac masses 
for the gauginos and Higgsinos. 
However, a special choice of charges is necessary to keep 
the Higgs boson light at tree level. 
Below the compactification scale, $R^{-1} \simeq 25$ TeV,
a conventional, logarithmic radiative EWSB occurs, with 
a spectrum which is similar to gauge mediation.
More generally, Scherk-Schwarz
supersymmetry breaking offers the prospect of a predictive superpartner
spectrum where radiative corrections are finite and dominated by the
compactification scale \cite{ADPQ}. This softness results because 
the supersymmetry breaking involves the global structure of the 
modes, and is therefore non-local. At distances beneath the compactification 
scale supersymmetry breaking effects are exponentially damped; this is quite 
unlike the case of supersymmetry in 4d, where radiative EWSB can originate 
from distances many orders of magnitude smaller than the weak scale.
Given this exciting result, it is
surprising that more explicit theories have not been constructed.

In this paper we construct a theory by combining Kaluza-Klein EWSB with
non-local supersymmetry breaking. We introduce a single compact
dimension of radius $R$ in which every particle of the standard model
propagates.\footnote{
Most studies of extra dimensions at the TeV scale have considered the
case that the Higgs bosons propagate in the bulk, while the quarks and
leptons do not \cite{BM1, BM2, PWM, DPQ2}. 
This is surprising since the orbifold construction
allows an elegant understanding of why the lightest bulk modes are
chiral -- a property which is crucial for matter but irrelevant for
the Higgs. A possible reason for this is the power law running of
couplings in 5d: with matter in the bulk the top Yukawa coupling grows
more rapidly and the gauge couplings become non-perturbative before
they unify. We insist that all standard model particles propagate in
the bulk, and find that there is an energy interval in which the 5d
theory is perturbative. Even with perturbative
power law unification of gauge couplings, the prediction for the weak
mixing angle is unreliable,
as it is quadratically sensitive to the physics at the cutoff.}
The bulk of the 5d theory is supersymmetric, so the KK towers have 
multiplets corresponding to two supersymmetries in 4d. Two orthogonal
reflection symmetries are imposed on the circle. One removes half of
the modes in the KK towers, so that the zero modes possess a single 
supersymmetry. This zero mode structure is chiral, as needed for
the quarks and leptons, but gives an electroweak gauge anomaly from the
Higgsino. However, the second reflection symmetry reduces the number
of supersymmetries to zero, yielding zero modes with precisely 
the states of the 1 Higgs doublet standard model.
Supersymmetry breaking by this second reflection can be viewed as 
a discrete version \cite{Dis-SS} of the Scherk-Schwarz mechanism.
Imposing one reflection symmetry to break one supersymmetry and obtain 
chiral zero modes has been widely used in the literature; the novel
feature of our theory is the imposition of two reflections to break
both supersymmetries.
The bulk is remarkably pristine, containing just four parameters: the
three gauge couplings and the compactification scale $R^{-1}$, which
sets the mass scale for the KK towers. Every standard model particle
is massless, and all the superpartners of the standard model have mass
$R^{-1}$. 

The physics of flavor occurs on branes not in the bulk.
The reflection symmetries allow supersymmetric Yukawa interactions to
be placed at their fixed points: for the up sector at the fixed point of one
reflection, and for the down and charged lepton sectors at the fixed
point of the second reflection. As the two reflection symmetries leave
different supersymmetries unbroken, these Yukawa couplings
involve the same Higgs doublet, even though the
underlying theory is highly supersymmetric. These brane Yukawa interactions
involve the usual 13 physical flavor parameters of the standard
model. As in the case of conventional supersymmetry, there is no
immediate progress in the understanding of flavor. Our theory is
described in detail in section \ref{sec:theory}.

The Higgs boson interacts with the entire KK top quark tower on the
branes containing the top quark Yukawa coupling, and Kaluza-Klein EWSB is
induced. However, there are no extra free parameters to describe the
resulting Higgs potential. Thus, the Fermi constant is used to
determine the compactification scale, and the Higgs boson mass is predicted.
The one-loop calculation of the compactification scale and the Higgs
boson mass is presented in section \ref{sec:higgs}. 
In section \ref{sec:sens} we show that the
Higgs boson mass has very little sensitivity to unknown physics at
short distances, although in the case of the compactification scale
there is some sensitivity to the ultraviolet (UV). We also study the 
$\rho$ parameter and find that it has quadratic sensitivity to the UV.

Ignoring brane interactions, all superpartners are degenerate with
mass $R^{-1}$. The large top Yukawa not only induces EWSB, but
significantly modifies the spectrum of the top squarks, the
neutralinos and the charginos. These masses are computed in section 
\ref{sec:lsp}, where we show that the LSP is a top squark. The collider 
phenomenology of the superpartners is briefly discussed, and we note that 
our theory possesses an anomaly free $U(1)_R$ symmetry. 
Conclusions are drawn in section \ref{sec:concl}.

\section{The Theory and the Tree-level Spectrum} \label{sec:theory}

We introduce a compact spatial dimension $y$, with a size of 
order an inverse TeV, and require that the 5d theory is supersymmetric. 
The standard model gauge bosons propagate in 5d and are contained
in 5d vector supermultiplets $(A^M, \lambda, \lambda', \sigma)$. The
5d vector field $A^M$ contains the standard model gauge boson from its
first 4 components, and a 4d scalar $A^5$. There are two gauginos,
$\lambda$ and $\lambda'$, reflecting the presence of two
supersymmetries on reduction to 4d, and a real scalar $\sigma$. It is
often convenient to view the theory as a $N=1$ 4d theory; the 5d
vector multiplet can be decomposed into a 4d vector supermultiplet
$V(A^\mu, \lambda)$, and a chiral multiplet in the adjoint
representation $\Sigma(\phi_\Sigma,\psi_\Sigma)$, where 
$\phi_\Sigma = (\sigma + iA^5) / \sqrt{2}$ and $\psi_\Sigma = \lambda'$.

The standard model matter and Higgs fields also propagate in 5d, and
are described by a set of hypermultiplets 
$(\Psi,\phi,\phi')_X$, where $\Psi$ is a Dirac fermion, and $\phi,\phi'$
are two complex scalars. 
These each decompose into two 4d $N=1$ chiral multiplets
$X(\phi_X,\psi_X)$ and $X^c(\phi^c_X, \psi^c_X)$, where $\Psi =
(\psi, \psi^{c \dagger})$ and $\phi' = \phi^{c \dagger}$.
Conjugated objects have conjugate transformations under the gauge
group, and X runs over three generations of matter, $Q,U,D,L,E$, and a
{\it single} Higgs $H$. 

The 5d theory contains a $SU(2)_R$ symmetry under which $(\lambda,
\lambda')$ and $(\phi, \phi')$ are doublets, while all other fields
are singlets. The two supersymmetries of the 4d theory are related by
$\lambda \leftrightarrow \lambda'$ and $\phi \leftrightarrow \phi'$
and play a crucial role in the construction of our theory. 
The 4d $N=1$ language, which we use from now on, hides both the
$SU(2)_R$ symmetry and the second supersymmetry.
The $N=1$ fields depend on $y$ as a continuous parameter; on 
compactification this leads to KK towers of $N=1$ fields.

\subsection{The $S^1/Z_2$ orbifold}

If the extra dimension is taken to be a circle, $S^1$, of radius $R$,
then each field has modes $e^{\pm i ny/R}$, $n=0,1,2,...$, and the
fermionic matter and Higgsino states are vectorial. To obtain chiral
matter, we first consider  restricting the space of the extra
dimension to the orbifold $S^1/Z_2$. A $Z_2$ symmetry is introduced
under which $y \rightarrow -y$ and all fields are either even or
odd, having modes
\begin{eqnarray}
&& +: \;\;\; \cos{n\,y \over R} \\
&& -: \;\;\; \sin{n\,y \over R}
\label{eq:z2modes}
\end{eqnarray}
respectively, so that only the even fields contain zero modes.
Since $A^\mu$ appears in the covariant derivative $D^\mu$ the
vector multiplet is even, $V_+$. On the other hand 
$\Im{\phi_\Sigma}$ appears in $D^5$, 
and therefore the chiral adjoint multiplet is odd,
$\Sigma_-$. The term in the superpotential $X D^5 X^c$ ensures that
$X$ and $X^c$ have opposite $Z_2$ charges. The conjugate label
is arbitrary, and without loss of generality we may choose 
$X_+, X^c_-$ --- the conjugate fields do not 
contain zero modes. These assignments are the same nomatter whether 
$X$ is matter or Higgs --- the orbifold only has one type of 
hypermultiplet and does not provide any distinction between matter 
and Higgs. The physical space of the orbifold is the
interval $0 < y < \pi R$, since once a field configuration is specified in
this region, it is fixed everywhere on the circle. The end points of
the interval are fixed points under the $y \rightarrow -y$ transformation.
The orbifolding has therefore broken $N=2$ to $N=1$ supersymmetry, 
and 5d Lorentz to 4d Lorentz symmetry at the boundary.

The symmetries of the bulk forbid interactions which break the flavor 
group $U(3)^5$ acting on $Q,U,D,L,E$. In particular,
bulk Yukawa interactions are forbidden by supersymmetry. However,
we can introduce interactions on the 4d subspaces of the orbifold fixed 
points, which we call brane interactions. Such interactions
necessarily violate 5d Lorentz symmetry and $N=2$ supersymmetry. 
However, we insist that they preserve the same spacetime symmetries
as the orbifold: $Z_2$, 4d Lorentz symmetry and $N=1$ supersymmetry. 
Note that we started with two supersymmetries, and we could have the 
orbifold and brane interactions break different supersymmetries. 
This would give a theory with supersymmetry completely broken in a 
hard way, and hence we require that the $Z_2$ orbifold and brane 
interactions leave one supersymmetry unbroken. 

The allowed brane superpotential depends on the hypercharge sign choice for
the Higgs chiral multiplet, $H$, which contains the zero mode. One
choice allows $QUH$, while the other allows $(QDH + LEH)$ --- we can get
masses for the up sector, or for the down and charged lepton sectors, but
not for both.\footnote{
Brane interactions involving $X^c$ fields are also 
allowed by gauge invariance, but all components of $X^c$ vanish on 
the orbifold fixed points, and hence these interactions also vanish.
$R$-parity violating interactions, such as $LLE$, are also allowed.}
One can take various views on this problem: perhaps it is
not a problem, but a useful zeroth-order approximation allowing an
understanding of why $m_t \gg m_b$, although the anomalies of the 
zero modes need to be cancelled. Alternatively, one could introduce
two Higgs hypermultiplets, so that both choices can be made. This will
lead to an minimal supersymmetric standard model (MSSM)-like 
4d theory at scales below $1/R$.  In this case, the extra
dimension has not addressed many of the issues familiar from the MSSM:
supersymmetry breaking and mediation, the distinction between matter
and Higgs and the origin of $R$ parity. In this
paper we explore a third alternative. 

We have seen that orbifolding with one $Z_2$ breaks one supersymmetry 
and allows one type of Yukawa coupling. 
The freedom of choice of the charge of $H$ turns out to be
equivalent to the choice of which supersymmetry is kept unbroken by 
the orbifolding. We find that if we orbifold twice, 
using two $Z_2\,$s, both supersymmetries
can be broken, and both types of Yukawa coupling are allowed.
The $Z_2\,$s have different fixed points so that the two supersymmetries 
are broken at different locations in the bulk, maintaining the softness 
of radiative corrections.

\subsection{The $S^1 / (Z_2 \times Z_2')$ orbifold}

\begin{figure}
\begin{center} 
\begin{picture}(160,180)(-80,-90)
  \CArc(0,0)(60,0, 360)
  \DashLine(-80,0)(80,0){2} \Text(-90,0)[r]{$A$}
  \DashLine(0,-80)(0,80){2} \Text(0,90)[b]{$A'$}
  \Vertex(-60,0){3} \Text(-65,10)[br]{$y=0$}
  \Vertex(0,-60){3} \Text(-5,-70)[ur]{$y=\frac{\pi}{2}R$}
  \Vertex(60,0){3}  \Text(65,-10)[ul]{$y=\pi R$}
  \Vertex(0,60){3}  \Text(5,70)[bl]{$y=-\frac{\pi}{2}R$}
\end{picture}
\caption{$S^1 / (Z_2 \times Z_2')$ orbifold in the fifth dimension.}
\label{Fig_space}
\end{center}
\end{figure}
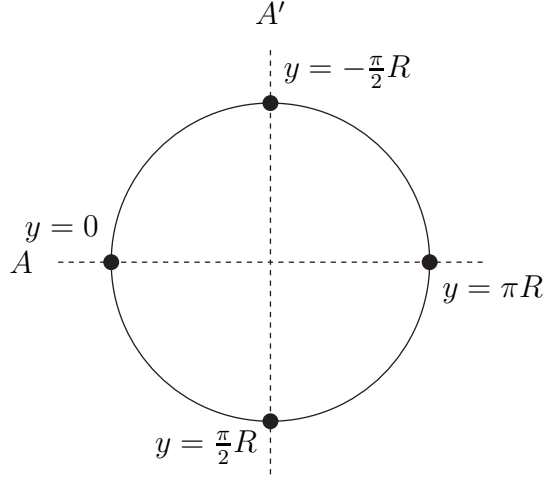

The $S^1 / (Z_2 \times Z_2')$ orbifold is constructed from the circle by
imposing two parities: $Z_2: \; y \rightarrow -y$ and  $Z_2': \;
y' \rightarrow -y'$, where $y' = y - \pi R /2$. These correspond to
reflections about the axes $A$ and $A'$ in Figure \ref{Fig_space}. 
The modes of the circle are now assembled into 4 types rather than 2, 
according to their $(Z_2,Z_2')$ quantum numbers:
\begin{eqnarray}
&& (+,+): \;\;\; \cos{2n\,y \over R} \label{eq:z2z2modes1}\\
&& (+,-): \;\;\; \cos{(2n+1)\,y \over R} \\
&& (-,+): \;\;\; \sin{(2n+1)\,y \over R} \\
&& (-,-): \;\;\; \sin{(2n+2)\,y \over R}
\label{eq:z2z2modes}
\end{eqnarray}
with $n=0,1,2,...$. Any component field will have just one type of mode,
according to its $Z_2 \times Z_2'$ assignment; only fields with (+,+)
assignment contain a zero mode.
The modes are completely specified over the circle once they are given 
on the interval $0 <y< \pi R/2$, which we choose to be the physical space.

There are two ways to interpret the quantum numbers of the two 
discrete $Z_2$ symmetries.  One way is the following.  The first
$Z_2$, which is the reflection $y \rightarrow -y$, leaves a
supersymmetry $S$ unbroken, and gives
$N=1$ multiplets with the usual $Z_2$ orbifold quantum
numbers discussed in the previous sub-section, such as $X_+$ and
$X^c_-$. Brane interactions, located at the fixed points at $y=0,\pi R$, 
should preserve $S$. For example, choosing $H$ to have positive
hypercharge allows the superpotential term $QUH$. When $Z_2'$ is 
introduced, the couplings of this interaction on the two branes at $y=0$ 
and $y= \pi R$ are constrained to be equal.

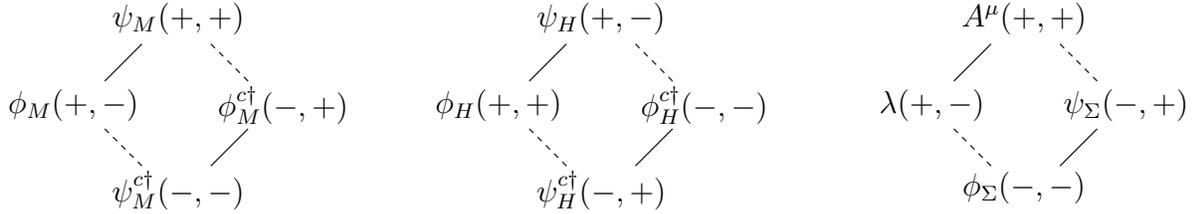
\begin{figure}
\begin{center} 
\begin{picture}(450,100)(-260,-20)
  \Text(-200,65)[b]{$\psi_{M} (+,+)$}
  \Text(-215,35)[r]{$\phi_{M} (+,-)$}
  \Text(-185,35)[l]{$\phi^{c\dagger}_{M} (-,+)$}
  \Text(-200,10)[t]{$\psi^{c\dagger}_{M} (-,-)$}
  \Line(-227,44)(-213,58) \DashLine(-173,44)(-187,58){2}
  \Line(-173,26)(-187,12) \DashLine(-227,26)(-213,12){2}
  \Text(-40,65)[b]{$\psi_{H} (+,-)$}
  \Text(-55,35)[r]{$\phi_{H} (+,+)$}
  \Text(-25,35)[l]{$\phi^{c\dagger}_{H} (-,-)$}
  \Text(-40,10)[t]{$\psi^{c\dagger}_{H} (-,+)$}
  \Line(-67,44)(-53,58) \DashLine(-13,44)(-27,58){2}
  \Line(-13,26)(-27,12) \DashLine(-67,26)(-53,12){2}
  \Text(120,65)[b]{$A^{\mu} (+,+)$}
  \Text(105,35)[r]{$\lambda (+,-)$}
  \Text(135,35)[l]{$\psi_{\Sigma} (-,+)$}
  \Text(120,10)[t]{$\phi_{\Sigma} (-,-)$}
  \Line(93,44)(107,58)  \DashLine(147,44)(133,58){2}
  \Line(147,26)(133,12) \DashLine(93,26)(107,12){2}
\end{picture}
\caption{Quantum numbers of the matter, Higgs and gauge 
multiplets under the two orbifoldings $y \rightarrow -y$ and 
$y' \rightarrow -y'$.}
\label{Fig_transf}
\end{center}
\end{figure}

How does $Z_2'$ act on these $N=1$ multiplets? This action is identified with
the $R$ parity, $R_P$, of this $S$ supersymmetry, i.e.
with $\theta \rightarrow - \theta$. For any hypermultiplet, $X$ can be
chosen to be $R_P$ even or odd --
there are now two types of hypermultiplet, one in which the zero mode
is a scalar, and the other where the zero mode is a fermion. Thus
supersymmetric theories on the  $S^1 / (Z_2 \times Z_2')$ orbifold
provide an inherent distinction between Higgs and matter multiplets
depending on whether their $Z_2$ and $Z_2'$ quantum numbers are the
same or different: $H(+,+), M(+,-)$. Invariance of the
interactions $H D^5 H^c$ and $M D^5 M^c$, then, determines the quantum
numbers for the conjugate multiplets $H^c(-,-), M^c(-,+)$. Covariance
of the gauge derivatives $D^\mu$ and $D^5$ determines the assignments
for the vector and chiral adjoint multiplets: $V(+,+), \Sigma(-,-)$. 
Making the usual superfield expansions, $H = \phi_H + \theta \psi_H$,
etc, gives the quantum numbers
for the components of the Higgs, matter and vector hypermultiplets
shown in Figure \ref{Fig_transf}. 

It is interesting to note that anomaly freedom of the low energy 
effective theory does not allow a single Higgs hypermultiplet 
with a circle reduced by a single $Z_2$ orbifolding. 
Two $Z_2$ orbifoldings are necessary to restore anomaly freedom. 
For matter hypermultiplets, however, fermionic zero
modes remain after orbifolding with both $Z_2\,$s --- anomalies cancel
between multiplets as in the standard model.

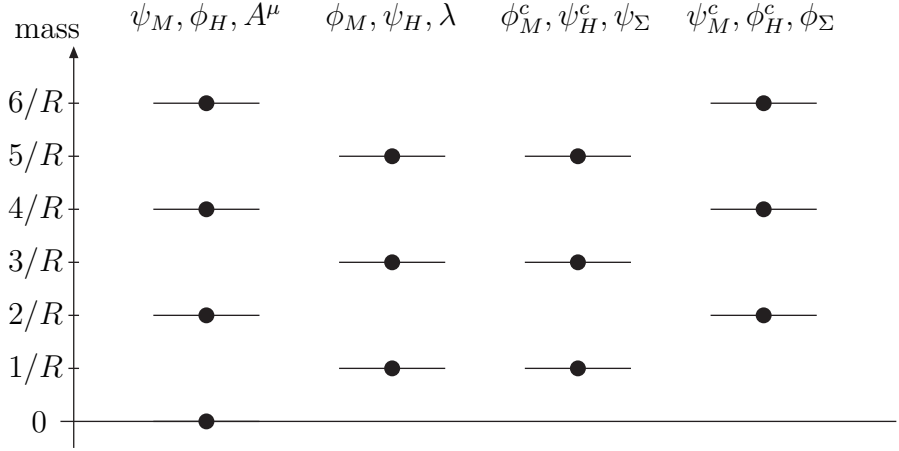
\begin{figure}
\begin{center} 
\begin{picture}(350,190)(-10,-25)
  \Line(5,0)(320,0)
  \LongArrow(10,-10)(10,140)
  \Text(0,145)[b]{mass}
  \Text(0,0)[r]{$0$}
  \Line(8,20)(12,20)    \Text(6,20)[r]{$1/R$}
  \Line(8,40)(12,40)    \Text(6,40)[r]{$2/R$}
  \Line(8,60)(12,60)    \Text(6,60)[r]{$3/R$}
  \Line(8,80)(12,80)    \Text(6,80)[r]{$4/R$}
  \Line(8,100)(12,100)  \Text(6,100)[r]{$5/R$}
  \Line(8,120)(12,120)  \Text(6,120)[r]{$6/R$}
  \Text(60,150)[b]{$\psi_{M}, \phi_{H}, A^{\mu}$}
  \Line(40,0)(80,0)      \Vertex(60,0){3}
  \Line(40,40)(80,40)    \Vertex(60,40){3}
  \Line(40,80)(80,80)    \Vertex(60,80){3}
  \Line(40,120)(80,120)  \Vertex(60,120){3}
  \Text(130,150)[b]{$\phi_{M}, \psi_{H}, \lambda$}
  \Line(110,20)(150,20)    \Vertex(130,20){3}
  \Line(110,60)(150,60)    \Vertex(130,60){3}
  \Line(110,100)(150,100)  \Vertex(130,100){3}
  \Text(200,150)[b]{$\phi^{c}_{M}, \psi^{c}_{H}, \psi_{\Sigma}$}
  \Line(180,20)(220,20)    \Vertex(200,20){3}
  \Line(180,60)(220,60)    \Vertex(200,60){3}
  \Line(180,100)(220,100)  \Vertex(200,100){3}
  \Text(270,150)[b]{$\psi^{c}_{M}, \phi^{c}_{H}, \phi_{\Sigma}$}
  \Line(250,40)(290,40)    \Vertex(270,40){3}
  \Line(250,80)(290,80)    \Vertex(270,80){3}
  \Line(250,120)(290,120)  \Vertex(270,120){3}
\end{picture}
\caption{Tree-level KK mass spectrum of the matter, Higgs and gauge 
multiplets.}
\label{Fig_spectrum}
\end{center}
\end{figure}

The second orbifold clearly
breaks supersymmetry, since the components of an $N=1$ superfield have
different modes. The tree level spectrum is shown in Figure 
\ref{Fig_spectrum}. The zero modes are in one-to-one correspondence 
with the particles of the standard model, and have KK
excitations with masses $2n/R$. The superpartners of standard model
particles, and the states obtained by an $SU(2)_R$ transformation
on these superpartners, have KK towers of states of mass
$(2n+1)/R$.
Finally, the conjugate quarks, conjugate Higgs and
$\phi_\Sigma$ have KK towers of mass $(2n+2)/R$.
At an arbitrary point in the bulk, supersymmetry is also broken by the
wavefunctions of the modes. At the orbifold fixed points at $y=0$ and
$y= \pi R/2$, the supersymmetry of the wavefunctions is restored,
except for the zero mode.

The second interpretation of the $(Z_2, Z_2')$ quantum numbers results
if we first orbifold the circle by $Z_2'$ about the axis $A'$ of
Figure \ref{Fig_space}. This produces the same orbifold discussed 
in the previous sub-section, $S^1/Z_2'$, with fixed points
rotated by $\pi /2$ to $y= \pm \pi R/2$.  
The crucial point is that the supersymmetry left unbroken by this 
orbifolding, $S'$, is not the
supersymmetry $S$, which is preserved by first orbifolding by $Z_2$
about axis $A$. The $S'$ multiplets are easily identified by grouping
together the $Z_2' = +$ and $Z_2' = -$ component fields of 
Figure \ref{Fig_transf}:
\begin{equation}
  M' \pmatrix{\phi_M^{c\,\dagger} \cr \psi_M} \;\;\;
  M^{c\,\prime} \pmatrix{\phi_M^\dagger \cr \psi_M^c} \;\;\;
  H' \pmatrix{\phi_H^{c\,\dagger} \cr \psi_H} \;\;\;
  H^{c\,\prime} \pmatrix{\phi_H^\dagger \cr \psi_H^c} \;\;\;
  V' \pmatrix{A^\mu \cr \psi_\Sigma} \;\;\;
  \Sigma' \pmatrix{\phi_\Sigma \cr \lambda}.
\label{eq:s'mult}
\end{equation}
The $S'$ chiral superfields are labelled by the left-handed fermions 
they contain, while the $S'$ vector multiplets are labelled by the 
bosonic components. The zero modes lie in $M', H^{c\,\prime}$ and $V'$. 
On comparing with the $S$ multiplets
\begin{equation}
  M \pmatrix{\phi_M \cr \psi_M} \;\;\;
  M^c \pmatrix{\phi_M^c \cr \psi_M^c} \;\;\;
  H \pmatrix{\phi_H \cr \psi_H} \;\;\;
  H^c \pmatrix{\phi_H^c \cr \psi_H^c} \;\;\;
  V \pmatrix{A^\mu \cr \lambda} \;\;\;
  \Sigma \pmatrix{\phi_\Sigma \cr \psi_\Sigma},
\label{eq:smult}
\end{equation}
one discovers that the transformation between $S$ and $S'$ is
accomplished by the interchanges 
$(\phi_X \leftrightarrow \phi_X^{c\,\dagger})$
and $(\lambda \leftrightarrow \psi_\Sigma)$, a discrete subgroup of
the $SU(2)_R$ symmetry. This interchanging between $S$ and $S'$
multiplets is easily visualized from the multiplet arrangements of
Figure \ref{Fig_transf}. The $(Z_2, Z_2')$ quantum numbers of both $S$ 
and $S'$ superfields are also easily read from this Figure, as they are 
always given by the quantum numbers of the boson.

The brane interactions at the $Z_2'$ fixed points $y= \pm \pi R/2$ are
$(Q' D' H^{c\,\prime} + L' E' H^{c\,\prime})$. Remarkably, since 
$H^{c\,\prime}$ contains the zero mode Higgs boson, these are precisely 
the interactions needed to give mass to down-type quarks and charged
leptons. Why is there not a complete symmetry under the interchange
$Z_2 \leftrightarrow Z_2'$ and $S \leftrightarrow S'$? The interchange
 $(\phi_X \leftrightarrow \phi_X^{c\,\dagger})$ results in the zero mode 
Higgs boson lying in chiral multiplets of opposite hypercharge in the two
cases, and this changes the form of the gauge-invariant brane interactions.

In this second viewpoint, $Z_2$ is identified as the $R$ parity, $R_P'$, 
of the supersymmetry $S'$, with $\theta' \rightarrow - \theta'$. Here
$\theta'$ is the superspace coordinate for $S'$, for example: $M'
= \phi_M^{c\,\dagger} + \theta' \psi_M$. This identification of $Z_2$
can be readily verified from the quantum numbers of Figure \ref{Fig_transf}.

The action of the discrete symmetries can be summarized by
\begin{eqnarray}
&& Z_2:  \;\;\; y \rightarrow -y, \quad (\theta' \rightarrow -\theta')_2, 
\nonumber \\
&& Z_2': \;\;\; y' \rightarrow -y', \quad (\theta \rightarrow -\theta)_1.
\label{eq:z2z2'}
\end{eqnarray}
The superspace coordinates $\theta$ and $\theta'$ are those of
different orthogonal $N=1$ supersymmetry subgroups --- they are not
the superspace coordinates of the full 5d theory.
The first viewpoint uses the $\theta$ transformation, while the
second viewpoint uses the $\theta'$ ones. In the MSSM, $R$ parity is 
imposed in an ad hoc fashion to avoid proton decay --- it is not 
inherent to the formulation of the theory. In our theory, $R$ parity 
becomes part of the orbifolding symmetry of (\ref{eq:z2z2'}), and is 
an unavoidable consequence of the basic formulation of the theory. 
It therefore comes as no surprise that, when the interactions of the 
low energy theory are derived, they are found to conserve baryon and 
lepton numbers at the renormalizable level.\footnote{
$R$-parity violating interactions could be introduced, if we allow 
the coupling constants to have the opposite signs on two branes 
$y=0$ and $\pi R$ ($y=\pm\pi R/2$).}

Our theory may be formulated as a Scherk-Schwarz theory as follows. 
Start with a $S^1/Z_2$ orbifold of radius $R'$, as described in the
previous sub-section. On imposing the condition
\begin{equation}
  \phi(y + 2 \pi R') = R_P\, \phi(y),
\label{SS-cond}
\end{equation} 
on the modes of any component field $\phi$, one discovers that the
allowed modes are precisely those of 
Eqs.~(\ref{eq:z2z2modes1} -- \ref{eq:z2z2modes}), with
$R=2 R'$. However, this formulation hides the symmetry between 
$Z_2$ and $Z_2'$ symmetries, as it stresses the role of $Z_2$ and 
$R_P = Z_2 Z_2'$.

To summarize: we have taken every particle of the standard model to
propagate in a compact dimension of size $R$. In the case that the 5d
theory is supersymmetric, and the compact space is the orbifold
$S^1/(Z_2 \times Z_2')$, the effective theory beneath the scale $1/R$
is non-supersymmetric and chiral, having the gauge and multiplet 
structure of the standard model. 
The most general brane interactions, consistent with $Z_2 \times Z_2'$
and the supersymmetries preserved by each separate orbifold, up to
cubic order, are
\begin{eqnarray}
&&  {1 \over 2} \left( \delta (y) + \delta (y - \pi R) \right) 
\int d^2 \theta \left( \lambda_U  QUH \right)  
\nonumber\\
&+& {1 \over 2} \left( \delta (y- \frac{\pi}{2} R) + 
\delta (y + \frac{\pi}{2} R) \right) 
\int d^2 \theta'  \left(
\lambda_D Q'D'H^{c\,\prime} + \lambda_E L'E' H^{c\,\prime} \right) 
+ {\rm h.c.} 
\label{eq:braneint}
\end{eqnarray}
Together these interactions break supersymmetry completely --- but the
locality of the operators results in the breaking being soft, even
though the low energy 4d theory contains $\psi_Q \psi_U \phi_H +
\psi_Q \psi_D \phi_H^\dagger$.
At scales below $1/R$, these give the Yukawa coupling matrices of the
standard model proportional to the matrices $\lambda_{U,D,E}$.
Hence, the low energy effective theory is {\it precisely} 
the standard model, with the Higgs potential {\it constrained} 
to have the tree-level form:
\begin{equation}
  V_{H,0} = {g^2 + g'^2 \over 8} |\phi_H|^4,
\label{eq:V0}
\end{equation}
where $g$ and $g'$ are the standard model $SU(2)$ and $U(1)$ gauge
couplings. The absence of any free parameters in the Higgs potential
at tree-level is striking. In the next section we calculate the
radiative contributions to the Higgs potential from interactions
involving the large top Yukawa coupling. We find that the effects of
virtual KK towers leads at one loop to finite corrections to the Higgs
potential involving the single parameter $R$. The Higgs mass-squared
is negative, the Fermi constant is used to determine $R$,
and the physical mass of the Higgs boson is predicted.

\section{The Higgs Boson Mass and the Compactification Scale} 
\label{sec:higgs}

In the previous section, we have investigated the tree-level structure 
of the model and found that the matter content of the massless sector 
is precisely that of the standard model.
In this section, we calculate the one-loop 
effective potential of the Higgs boson coming from 
KK towers of the quark hypermultiplets through the top Yukawa coupling.
We find that the Higgs boson receives a negative mass-squared and 
EWSB is radiatively triggered.
Furthermore, the effective potential is 
finite and depends only on the top Yukawa 
coupling $y_t$ and the compactification radius $R$.
Thus, demanding the VEV of the Higgs field to be $175~{\rm GeV}$, 
we can calculate the value of the physical 
Higgs-boson mass and the compactification radius $R$, which  
determines the masses for the superpartners and the KK excitations.

\subsection{The Higgs mass squared}

Before computing the one-loop effective potential, we first calculate 
diagrammatically the mass-squared for the Higgs doublet by making 
a KK decomposition of the original 5d theory.
The interactions between the Higgs boson and the KK modes of the quark 
fields are read off by expanding the brane interaction given in 
Eq.~(\ref{eq:braneint}).
After eliminating the auxiliary $F$ fields, the relevant terms are
\begin{eqnarray}
  S_{\rm int} &=&
  \int d^4 x \Biggl[
    \sum_{k,l=0}^{\infty}
    \Bigl(\frac{f_t}{\sqrt{2}} m_{\phi^c,k} 
    \eta^{F}_k \eta^{\phi}_l
    \phi^{c\,\dagger}_{Q,k} \phi_{U,l} \phi_H
  + \frac{f_t}{\sqrt{2}} m_{\phi^c,k} 
    \eta^{F}_k \eta^{\phi}_l 
    \phi^{c\,\dagger}_{U,k} \phi_{Q,l} \phi_H
  + {\rm h.c.} \Bigr)
\nonumber\\
&&- \sum_{k,l,m=0}^{\infty}
    \Bigl(\frac{f_t^2}{2} 
    \eta^{\phi}_k \eta^{\phi}_l (\eta^{F}_m)^2
    \phi^{\dagger}_{Q,k} \phi_{Q,l} \phi^{\dagger}_H \phi_H
  + \frac{f_t^2}{2}
    \eta^{\phi}_k \eta^{\phi}_l (\eta^{F}_m)^2
    \phi^{\dagger}_{U,k} \phi_{U,l} \phi^{\dagger}_H \phi_H \Bigr)
\nonumber\\
&&- \sum_{k,l=0}^{\infty}
    \Bigl(\frac{f_t}{\sqrt{2}} 
    \eta^{\psi}_k \eta^{\psi}_l
    \psi_{Q,k} \psi_{U,l} \phi_H 
  + {\rm h.c.} \Bigr) \Biggr],
\label{bare-Lag}
\end{eqnarray}
where $\phi_H$ is the zero-mode Higgs boson; 
$\phi_{M,k}$, $\psi_{M,k}$, $\phi^c_{M,k}$ and $\psi^c_{M,k}$ 
$(M=Q,U)$ represent the $k$-th KK modes of the component fields 
in the hypermultiplet $M$ (see Figure \ref{Fig_transf}).
The dimensionless coupling $f_t$ is defined by 
$f_t \equiv (\lambda_U)_{33} / (\pi R)^{3/2}$; 
$\eta^{\phi}_k$, $\eta^{\psi}_k$ and $\eta^{F}_k$ are the values 
of the wavefunctions at $y = 0$ for the $\phi_{M,k}$, $\psi_{M,k}$ 
and $F_{M,k}$ fields, respectively.

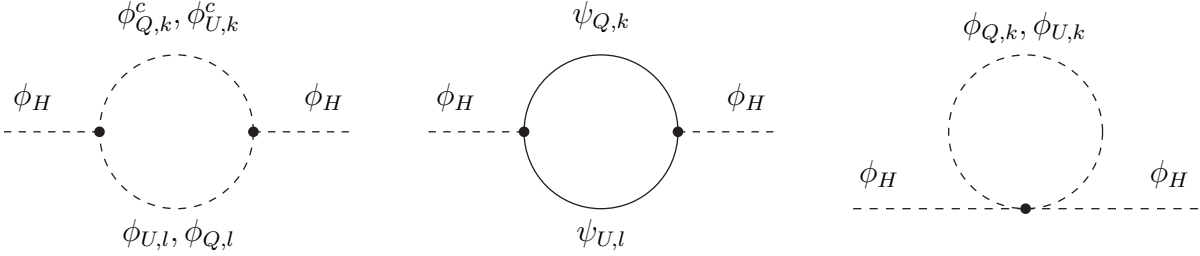
\begin{figure}
\begin{center} 
\begin{picture}(450,110)(-265,-45)
  \DashLine(-265,10)(-229,10){3} \Vertex(-229,10){2}
  \Text(-255,20)[b]{$\phi_H$}
  \DashCArc(-200,10)(29,0,360){3}
  \Text(-200,51)[b]{$\phi^c_{Q,k}, \phi^c_{U,k}$}
  \Text(-200,-23)[t]{$\phi_{U,l}, \phi_{Q,l}$}
  \DashLine(-171,10)(-135,10){3} \Vertex(-171,10){2}
  \Text(-145,20)[b]{$\phi_H$}
  \DashLine(-105,10)(-69,10){3} \Vertex(-69,10){2}
  \Text(-95,20)[b]{$\phi_H$}
  \CArc(-40,10)(29,0,360)
  \Text(-40,51)[b]{$\psi_{Q,k}$}
  \Text(-40,-23)[t]{$\psi_{U,l}$}
  \DashLine(-11,10)(25,10){3} \Vertex(-11,10){2}
  \Text(15,20)[b]{$\phi_H$}
  \DashLine(55,-19)(120,-19){3} \Text(65,-9)[b]{$\phi_H$}
  \DashCArc(120,10)(29,0,360){3} \Vertex(120,-19){2}
  \Text(120,47)[b]{$\phi_{Q,k}, \phi_{U,k}$}
  \DashLine(120,-19)(185,-19){3} \Text(175,-9)[b]{$\phi_H$}
\end{picture}
\caption{One-loop diagrams contributing to the mass squared 
of the Higgs boson.}
\label{Fig_RC-Higgs}
\end{center}
\end{figure}

The Higgs boson mass $m_{\phi_H}$ is generated at the one-loop level 
via loops of KK towers of the $Q$ and $U$ multiplets.
There are three diagrams, as shown in Figure \ref{Fig_RC-Higgs}, 
giving the contributions
\begin{eqnarray}
-i\, m_{\phi_H}^2 &=& 
  N_c f_t^2 \sum_{k,l=0}^{\infty} \int \frac{d^4 p}{(2\pi)^4} 
\nonumber\\
&& \times \left\{
    \frac{(\eta^{F}_k)^2 (\eta^{\phi}_l)^2 m_{\phi^c,k}^2}
    {(p^2-m_{\phi^c,k}^2)(p^2-m_{\phi,l}^2)}
  - \frac{(\eta^{\psi}_k)^2 (\eta^{\psi}_l)^2 p^2}
    {(p^2-m_{\psi,k}^2)(p^2-m_{\psi,l}^2)}
  + \frac{(\eta^{\phi}_k)^2 (\eta^{F}_l)^2}
    {(p^2-m_{\phi,k}^2)} \right\}.
\label{Higgs-gen}
\end{eqnarray}
Thus, substituting the KK masses and wavefunctions obtained in the 
previous section,
\begin{eqnarray}
  \left\{ \begin{array}{l}
    m_{\psi,k} = \frac{2k}{R} \equiv m_k \\ 
    m_{\phi,k} = m_{\phi^c,k} = \frac{2k+1}{R} \equiv \tilde{m}_k, \\
  \end{array} \right. \qquad
  \left\{ \begin{array}{l}
    \eta^{\phi}_k = \eta^{F}_k = 1 \\ 
    \eta^{\psi}_k = (\frac{1}{\sqrt{2}})^{\delta_{k,0}} 
    \equiv \eta_k, \\
  \end{array} \right.
\label{KK-param}
\end{eqnarray}
into Eq.~(\ref{Higgs-gen}), we obtain the one-loop induced Higgs boson 
mass-squared in the present model.

Performing a Wick rotation to Euclidian momentum space $p_E$, 
and changing to the variable $x = p_E R$, gives
\begin{eqnarray}
-i\, m_{\phi_H}^2 &=& 
   i N_c f_t^2 \sum_{k,l=0}^{\infty}
    \int \frac{d^4 p_E}{(2\pi)^4} 
\nonumber\\
&& \times \left\{
    \frac{\tilde{m}_k^2}{(p_E^2+\tilde{m}_k^2)(p_E^2+\tilde{m}_l^2)}
  + \frac{\eta_k^2 \eta_l^2\, p_E^2}{(p_E^2+m_k^2)(p_E^2+m_l^2)}
  - \frac{1}{(p_E^2+\tilde{m}_k^2)} \right\}
\\
&=& \frac{i N_c f_t^2}{R^2} 
    \int \frac{d^4 x}{(2\pi)^4} x^2 
\nonumber\\
&&  \times
    \sum_{k,l=0}^{\infty}
    \Biggl[ \frac{\eta_k^2 \eta_l^2}{(x^2+(2k)^2)(x^2+(2l)^2)} 
    - \frac{1}{(x^2+(2k+1)^2)(x^2+(2l+1)^2)} \Biggr].
\label{sum-mass}
\end{eqnarray}
In this expression, we first sum over the tower of KK states 
and then perform the momentum integral.
The resulting Higgs mass-squared is
\begin{eqnarray}
  m_{\phi_H}^2 &=&
  - \frac{N_c f_t^2}{128R^2} \int_0^{\infty}dx\, x^3 
  \left\{ \coth^2 \left[\frac{\pi x}{2}\right] 
  - \tanh^2 \left[\frac{\pi x}{2}\right] \right\}
\\
  &=& - \frac{21\,\zeta(3)}{64 \pi^4} \frac{N_c f_t^2}{R^2},
\label{Higgs-mass-1}
\end{eqnarray}
where $\zeta(x)$ is the Riemann's zeta function.
We find that the radiative correction $m_{\phi_H}^2$ is negative, 
so that EWSB is indeed triggered by loops involving the top KK towers.
Furthermore, the result is finite and UV insensitive, since the 
momentum integral is exponentially cut off at $p_E \sim R^{-1}$; 
99.99\% (99\%) of the integral comes from the region 
$p_E \lesssim 5/R$ ($p_E \lesssim 3/R$).
This extreme softness arises because the geometrical separation between 
the two orbifold fixed points in the extra dimension acts as a 
point-splitting regularization.
We can also rewrite Eq.~(\ref{Higgs-mass-1}) by using the 4d 
top Yukawa coupling $y_t = f_t / 2^{3/2}$ as
\begin{eqnarray}
  m_{\phi_H}^2 &=&
  - \frac{21\,\zeta(3)}{8 \pi^4} \frac{N_c\, y_t^2}{R^2}. 
\label{Higgs-mass} 
\end{eqnarray}
It is interesting to note that a similar calculation for the 
squark mass-squared gives a vanishing result, $m_{\phi_M}^2 = 0$.

\subsection{The effective potential}

We would like to compute the Higgs potential with sufficient accuracy
that the minimization leads to a determination of the compactification
scale and the Higgs mass to better than 10\%. Balancing the negative Higgs
mass squared, given above, against the tree-level gauge quartic Higgs
interaction of Eq.~(\ref{eq:V0}) is not sufficient. In fact, it is not 
even sufficient to include the quartic interaction obtained from
integrating out the top KK tower, as can be seen from the following
simple argument. The only dimensionful parameter of our theory is $R$,
so that this must set the scale of the effective potential. The
contribution  obtained by integrating out the top KK tower at one loop
involves a factor of the top Yukawa coupling $y_t$ for each external  
Higgs field. Hence the one-loop top KK tower contribution takes the form
$V_t = f(x)/R^4$, where $x = y_t^2 R^2 \phi_H^\dagger
\phi_H$. Expanding $f(x) = Ax + Bx^2 + C x^2 \ln x + D x^3 + ...$, all
the coefficients, $A,B,C, ...$ arise at one loop and are expected to
be comparable. Truncation at finite order is therefore unreliable. 

The one-loop, all-orders, effective potential from integrating out the 
top KK tower is
\begin{equation}
  V_t(H) = {1 \over 2} \mbox{Tr} \int {d^4 p \over (2 \pi)^4} 
  \sum_{k= - \infty}^{+ \infty}
  \ln \left( {p^2 + m_{B_k}^2(H) \over p^2 + m_{F_k}^2(H)} \right),
\label{eq:Veff}
\end{equation}
where $H = |\phi_H|$ and the trace is taken over all states 
of the top hypermultiplet of a given
$k$, giving a factor $4 N_c$. Clearly we only need the field dependent
eigenvalues of the boson and fermion mass matrices $m_{B_k}$ and 
$m_{F_k}$. The top Yukawa
coupling on the brane leads to a mixing of the tree level modes
Eqs.~(\ref{eq:z2z2modes1} -- \ref{eq:z2z2modes}) of the top states. 
For example, to leading order in $\left\langle \phi_H \right\rangle R$,
the effective theory below $2/R$ contains mass matrices for the
squarks of the form\footnote{
Integrating out the tower of $\phi^c_{M,k}$ $(k=1,2,\cdots)$ 
generates the term 
$\sum_{k=1}^{\infty} (f_t^2/2) \phi^{\dagger}_{M,0} \phi_{M,0} 
\phi^{\dagger}_{H} \phi_{H}$ in the low-energy Lagrangian, which 
cancels infinities present in the bare Lagrangian Eq.~(\ref{bare-Lag}), 
$-\sum_{k=0}^{\infty} (f_t^2/2) \phi^{\dagger}_{M,0} \phi_{M,0} 
\phi^{\dagger}_{H} \phi_{H}$.}
\begin{eqnarray}
  -{\cal L}_{\rm mass} = 
  \pmatrix{
     \phi^{\dagger}_{Q,0} & \phi^{c\,\dagger}_{U,0} \cr}
  \pmatrix{
     \left( \frac{1}{R} \right)^2 + 4 m_t^2  & 
     - 2 m_t \frac{1}{R}  \cr
     - 2 m_t \frac{1}{R} &
     \left( \frac{1}{R} \right)^2 \cr}
  \pmatrix{
     \phi_{Q,0}  \cr  \phi^c_{U,0}  \cr}
  + \left( Q \leftrightarrow U \right).
\label{mass_1}
\end{eqnarray}
However, this leading order matrix does not include exactly the
effects of mixing between these states and the heavier states.
We find the exact tree-level eigenvalue conditions to be
\begin{equation}
  \tan^2 \left( {\pi R\, m_{F_k} \over 2} \right) 
  =  {(\pi y_t R H)^2 \over 4}, 
\label{eq:fermcond}
\end{equation}
and
\begin{equation}
  \cot^2 \left( {\pi R\, m_{B_k} \over 2} \right) 
  =  {(\pi y_t R H)^2 \over 4},
\label{eq:boscond}
\end{equation}
giving eigenvalues
\begin{equation}
  m_{B_k}(H) = { 2k+1 \over R} \pm m_t(H)
  \qquad (k = 0,1,2, \cdots),
\label{stopmasses}
\end{equation}
and
\begin{equation}
  m_{F_k}(H) = { 2k \over R} \pm m_t(H)
  \qquad (k = 1,2,3 \cdots).
\label{topmasses}
\end{equation}
The zero-th order degeneracy of each level is
split by $2m_t$. At each mass eigenvalue there is a single Dirac
fermion, or two complex scalars.
There is also a single $k=0$ fermion mode, the
top quark, with mass
\begin{equation}
  m_t(H) = {2 \over \pi R} \arctan \left( {\pi y_t R H \over 2} \right).
\label{topmass}
\end{equation}
As $H$ varies from 0 to $\infty$, $m_t(H)$ grows from 0 to a
maximum value of $1/R$.  
This dependence of the top quark mass on the electroweak field is 
quite unlike the standard model. For small $y_t$ it reduces to the 
standard model result $m_t = y_t H$, but, for the observed value of 
the top mass, the effect of mixings with the heavier KK modes is important.
Increasing $H$ leads to a larger
splitting of each level, but there is no level crossing.
These eigenvalues can be used in Eq.~(\ref{eq:Veff}), with 
$-\infty < k < +\infty$, by choosing the positive sign in 
Eqs.~(\ref{stopmasses}, \ref{topmasses}),
and, using calculational techniques from Ref.~\cite{ADPQ}, we find
\begin{equation}
  V_t(H) = {6 N_c \over \pi^6 R^4} 
  \sum_{k=0}^\infty {\cos[(2k+1) \pi R\, m_t(H)] \over (2k+1)^5}.
\label{Vt}
\end{equation}
The denominator ensures that the higher terms
in this series are rapidly suppressed.
Taylor expanding around $H=0$, the quadratic term in this potential 
reproduces the mass-squared of Eq.~(\ref{Higgs-mass}) obtained by 
diagrammatic calculation.

\begin{figure}
\centerline{\epsfxsize=11cm \epsfbox{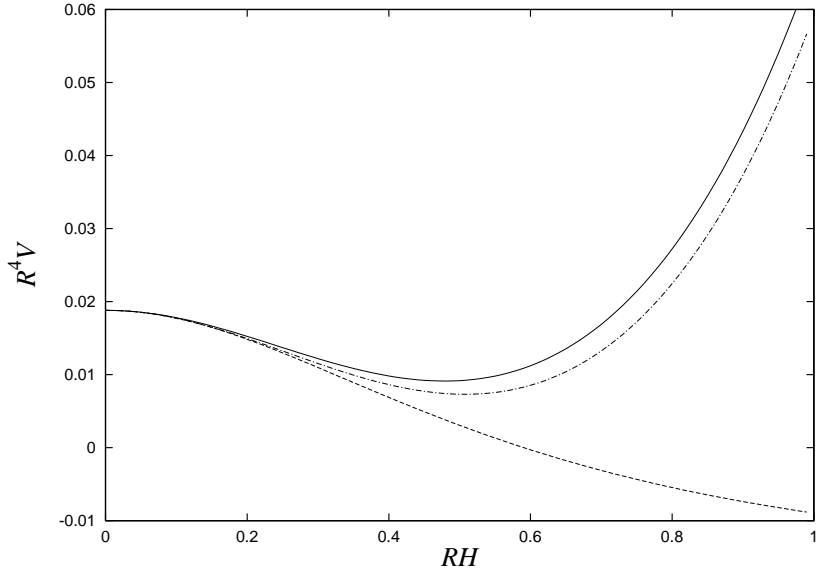}}
\caption{The Higgs potential $V$ normalized by the compactification 
radius $R$, as a function of $R H \equiv R |\phi_H|$.
Dashed, dot-dashed and solid lines represent $V_t$, 
$V_H \equiv V_{H,0} + V_t$ and $V_H$ with a gauge quadratic term, 
respectively.}
\label{Fig_pot}
\end{figure}

The potential $V_t$ is a monotonically decreasing function of
$H$, as shown in Figure \ref{Fig_pot}. Runaway behaviour is prevented 
by the tree-level gauge potential $V_{H,0}$ of Eq.~(\ref{eq:V0}), 
so that the combined potential $V_H \equiv V_{H,0} + V_t$ has the minimum 
shown in Figure \ref{Fig_pot}, given by the minimization condition
\begin{equation}
  {1 \over R} = \left( {\pi^6 \over 18} \right)^{1 \over 4}
  (M_Z v)^{1 \over 2}  \simeq 341 \; \mbox{GeV},
\label{eq:minimiz}
\end{equation}
where $v = \left\langle H \right\rangle$. More precisely, the
right-hand side should be multiplied by the factor $\xi^{-1/4}$ where
\begin{equation}
  \xi = \sin [\pi R\, m_t] \sum_{k=0}^\infty {\sin[(2k+1)\pi R\, m_t]
      \over (2k+1)^4}.
\label{eq:xi}
\end{equation}
Accidentally, the minimum occurs at a value of $R$ such that $R\, m_t$
is very close to 1/2: $\sin [\pi R\, m_t]$ is unity to better than 1\%, 
and the deviation of $\xi$ from unity only affects $1/R$ at the level
of about 1 GeV.

In the last section we stressed that compactification of our 5d theory 
led to a low energy effective theory which was the standard model with
the Higgs potential constrained to have the tree-level form of
$V_{H,0}$. In this section, we have discovered that EWSB in the low
energy effective theory is quite unlike that of the standard model,
because the relevant Higgs potential involves interactions to all orders 
in the Higgs field, not just the quadratic and quartic terms. For example,
the Higgs VEV is not given by the familiar form $v^2 =
-m_{\phi_H}^2/\lambda$ where $\lambda$ is the quartic coupling. If
this were true we would find $v^2 \propto y_t^2 R^{-2}/(g^2+g'^2)$.
Minimization of the all orders potential has given a different dependence:
\begin{equation}
  v^4 = {1 \over (g^2 + g'^2)} {L \over R^4},
\label{eq:v}
\end{equation}
where $L$ is the loop factor giving the relative size of tree and
one-loop terms in the minimization condition $L=36/\pi^6$.
The scale of the VEV is still set by $R^{-1}$, and runaway is
prevented by the non-zero value of the gauge couplings, but, because
of the accident mentioned above, there is essentially no sensitivity to the
uncertainty in the experimental value of the top quark mass.

Given that $R^{-1}$ is the only scale in the problem, why have the
superpartners not yet been discovered? The origin of the difference
between $R^{-1}$ and $v=175$ GeV can be seen from Eq.~(\ref{eq:v}). 
The weakness of the gauge couplings actually increase $v$ above the
compactification scale by 16\%. However, the loop factor $L^{1/4}$
increases $R^{-1}$ above $v$ by a factor of 2.3. The weak gauge bosons
are lighter than the superpartners because EWSB is driven only at one 
loop. Nevertheless, because it is the fourth root of $L$ that appears
in $v$, the superpartner masses are not far above the weak boson
masses.

Expanding $V_H$ about the minimum gives the physical Higgs mass
\begin{equation}
  m_H = \sqrt{2} M_Z (1 - {1 \over 4} \cos [\pi R\, m_t]) 
    \simeq 127 \; \mbox{GeV},
\label{eq:mH}
\end{equation}
where omitted terms in the expansion affect the Higgs mass by less
than 1 GeV, and we have used a $\overline{\rm MS}$ top quark mass of 
166 GeV in evaluating $\cos [\pi R\, m_t] \simeq 0.06$. 
An uncertainty of 5 GeV in the experimental value of the top quark mass 
translates into a 2 GeV uncertainty on the Higgs boson mass.

\subsection{Predictions and uncertainties}

A correction to the effective potential of the Higgs boson
comes from one-loop radiative corrections from the KK tower of the
$SU(2) \times U(1)$ gauge multiplets. Here we
study only the contribution to the quadratic term in the potential
\cite{ADPQ}
\begin{eqnarray}
  \Delta  m_{\phi_H}^2 =
    \frac{21\,\zeta(3)}{16 \pi^4} 
    \frac{g^2 + g'^2/3}{R^2}.
\end{eqnarray}
The effect of including this quadratic term is shown in 
Figure \ref{Fig_pot}.
This increases $R^{-1}$ by 3\%.

The remaining uncertainties are the electroweak corrections to the 
higher terms in the effective potential and the two-loop top
contributions to the effective potential, which we estimate at 1\% and
6\% respectively. We have calculated the induced VEVs for the higher KK 
modes of the Higgs boson, and found these effects to be at the 1\% level. 
Hence the compactification scale is
\begin{eqnarray}
  \frac{1}{R} = 352 \pm 20~{\rm GeV},
\label{comp-scale}
\end{eqnarray}
giving superpartner masses of $\simeq 350~{\rm GeV}$ 
and masses for the first KK excitations of $\simeq 700~{\rm GeV}$.
The physical Higgs boson mass is 
\begin{eqnarray}
  m_H = 127 \pm 8 \pm 2~{\rm GeV}.
\label{phys-mass}
\end{eqnarray}
In these predictions, the first uncertainty is a combined theoretical 
uncertainty from the effects discussed above, while the second 
uncertainty in $m_H$ follows from the experimental uncertainty of 
$\pm 5$ GeV in the measured value of the top quark mass. 

In the above calculation we performed summations over an infinite tower 
of KK modes. However, as discussed in the next section, our theory becomes
strongly coupled at energies of about $5/R$ and must be cutoff in some
way.  Are we correct to use the exact mass eigenvalues of
Eqs.~(\ref{stopmasses}, \ref{topmasses}) rather than the eigenvalues of
mass matrices involving just the lower modes?  Equivalently: what is the
sensitivity of our results to the details of the UV cutoff?  One may 
worry that there is a large sensitivity as each term in the sum of
Eq.~(\ref{sum-mass}) is separately quadratically divergent.  The crucial 
point is that we require our cutoff to preserve supersymmetry.  Since
supersymmetry is broken non-locally at the scale $R^{-1}$, the cutoff 
must preserve the cancellations which occur in the naive KK summation.
This is non trivial, since a quadratic divergence still appears if 
the sums in Eq.~(\ref{sum-mass}) are terminated at some finite value, 
even keeping an equal number of bosonic and fermionic states.
Thus we assume that the infinite sum over the KK modes gives the correct 
regularization of the theory; deviations from this ``KK regularization'' 
must be extremely small.

Finally, we note that in the above calculation it has been assumed 
that the KK towers of the quark multiplets have masses quantized 
precisely in units of $1/R$.
In general, however, the KK excitations receive wavefunction 
renormalizations due to brane interactions and their masses 
are renormalized.
This can cause a shift of the compactification radius 
by as much as a few tens of percent, which thereby acquires some 
sensitivity to unknown UV physics.
Remarkably, the Higgs mass prediction is much less sensitive to 
these wavefunction renormalizations, so that even in this case 
the tightness of Eq.~(\ref{phys-mass}) is maintained.
This issue is discussed in the next section.

\section{Sensitivity to Physics above the Compactification Scale} 
\label{sec:sens}

In the previous section we computed one-loop radiative corrections to 
the Higgs potential, yielding a prediction of both the Higgs boson 
mass and the compactification scale $1/R$, which determines the masses 
of the superpartners and the KK excitations. 
The sum over KK modes softened the usual logarithmic divergence of 
supersymmetry to give a completely finite result, suggesting that 
the physics of EWSB is really governed by the energy scale $1/R$, 
and has little sensitivity to whatever physics occurs 
at much higher energies.  In this section we demonstrate that the
compactification scale does have some sensitivity to
physics in the UV, and we estimate this uncertainty. 
Remarkably, the Higgs mass is much less sensitive to 
these effects, so that the precise prediction of
Eq.~(\ref{phys-mass}) is still expected to hold.

\subsection{Constraints from precision electroweak data}

For many theories in which the standard model gauge interactions
propagate in a fifth dimension, precision
electroweak data place a bound on the radius of order 
$R^{-1} \gtrsim 3~{\rm TeV}$ \cite{PWM, DPQ2}.
These stringent bounds apply when matter or Higgs fields are located on
a brane, and arise because the interactions of brane fields with the bulk
gauge bosons violate momentum conservation in the extra dimension.
For example, the kinetic term $\delta(y) Q^\dagger e^{gV} Q$ introduces
interactions between the zero mode quarks and the excited modes of the
gauge bosons: $\bar{q}_0 \gamma_\mu A^\mu_k q_0$. This allows both the
production of single gauge KK modes, $A^\mu_k$, and the generation of four
zero-mode fermion operators from their virtual exchange. Similar
effects result for brane leptons, while the situation for the brane Higgs
is particularly dangerous. The operator $\delta(y) H^\dagger e^{gV} H$
causes mass mixing between the $W$ and $Z$ zero modes and their
excited modes, and also induces a VEV for the weak triplet scalar in
the $SU(2)$ chiral adjoint $\Sigma$ \cite{DPQ2}. 
All these effects are absent to leading order in
our theory because both matter and Higgs propagate in the bulk. All
interactions, apart from the Yukawa interactions, conserve momentum in
the fifth dimension, so that the effects that could produce the stringent
bounds do not occur. Non-zero mode states can only be pair produced,
and, for states of mass $1/R \simeq 350$ GeV, this 
will require further runs of the Tevatron and the LHC.

At one loop, the top Yukawa interaction generates brane kinetic terms
for $Q_3,U_3$ and $H$. The contribution from short distances $(<R)$ is
expected to be supersymmetric:
\begin{equation}
  {1 \over 2} \left( \delta (y) + \delta (y - \pi R) \right)
  \int d^4 \theta \left( Z_{Q_3} Q_3^\dagger e^{gV} Q_3
  + Z_{U_3} U_3^\dagger e^{gV} U_3 + Z_H H^\dagger e^{gV} H \right).
\label{eq:braneke}
\end{equation}
Even though $Q_3,U_3$ and $H$ propagate in the bulk with kinetic terms 
$ Q_3^\dagger e^{gV} Q_3$, etc,  the brane kinetic interactions
violate momentum conservation in the $y$ direction, leading to terms
linear in KK modes. We define a dimensionless brane $Z$ factor $z_X = 
Z_X/ (2 \pi R)$, so that, in the 4d theory, the
zero modes have kinetic terms with coefficient $1+z_X$. 
For $1/R$ in the region of 350 GeV, experiments require 
$|z_H| \lesssim 0.2$.
As we discuss below, $z_H$ is scale dependent --- this bound applies 
at the scale $1/R$.  Limits on $z$ for the light quarks and leptons, 
from tree-level contributions of electroweak gauge bosons to 
precision electroweak observables, are also
about 0.2. The experimental constraints on
$z_{Q_3}$ and $z_{U_3}$ for the third generation quarks are much
milder. We take this limit on the Higgs $Z$ factor to imply that it
is also reasonable for these quark $Z$ factors to be less than 0.2.
This is strengthened by the estimates provided below, suggesting values of
$z_{Q_3}$ and $z_{U_3}$ in the range of about 0.1.

What are the effects of these $Z$ factors on the results of the calculation 
of the previous section? 
The $Z$ factor for the Higgs field does not affect the results at all. It is
removed by going to canonical normalization before the calculation is
begun, and affects only the relation between the 4d and 5d top coupling.

To find the effect of the $Z$ factors for $Q_3$ or $U_3$ we proceed as
follows.  We make a KK mode expansion and rescale the fields of the
equivalent 4d theory to obtain canonical kinetic energy. We study the
mass matrices for the fermion and boson KK modes to linear order in
$z_{Q,U}$, and find that the eigenvalue conditions have changed from
Eqs.~(\ref{eq:fermcond}, \ref{eq:boscond}) to
\begin{equation}
  \tan \left( {\pi R\, m_{F_k} \over 2(1-z_Q)} \right) 
  \tan \left( {\pi R\, m_{F_k} \over 2(1-z_U)} \right) 
  =  {(\pi R y_t v)^2 \over 4 (1- z_Q) (1- z_U)}, 
\label{eq:fermcondz}
\end{equation}
and
\begin{equation}
  \cot \left( {\pi R\, m_{B_k} \over 2(1-z_Q)} \right) 
  \cot \left( {\pi R\, m_{B_k} \over 2(1-z_U)} \right) 
  =  {(\pi R y_t v)^2 \over 4 (1- z_Q) (1- z_U)}, 
\label{eq:boscondz}
\end{equation}
which, to linear order in $z_{Q,U}$, reduces to
Eqs.~(\ref{eq:fermcond}, \ref{eq:boscond}) with the replacement
$1/R \rightarrow (1-\bar{z})/R$, where $\bar{z} = (z_Q + z_U)/2$.
With this replacement, Eqs.~(\ref{stopmasses} -- \ref{eq:mH})
all apply, so that it is $ (1-\bar{z})/R$ which is determined 
by the minimization of the Higgs potential to be 352 GeV. 
The numerical predictions for the top squark masses and Higgs 
are unaltered, while the other superpartner masses are
\begin{equation}
  \frac{1}{R} \simeq (1+ \bar{z}) (352 \pm 20)~{\rm GeV}.
\label{eq:compactscale}
\end{equation}
For $|z_{Q,U}| < 0.2$, the uncertainty on $1/R$ from $Z$ factors is 20\%.
The Higgs and top squark masses are affected only at quadratic order
in $z_{Q,U}$, which is at the percent level.

\subsection{Power law running}

In the 5d theory the gauge and Yukawa couplings display power law
running behaviour. The $Z$ factor for the vector field, and therefore
for the gauge coupling, is linearly divergent. From the 4d viewpoint,
the KK modes imply that the one-loop beta function coefficient 
at scale $E$ is proportional to $ER$, the number of KK modes lighter 
than $E$. For the chiral fields $Q_3, U_3$ and $H$, the 5d interaction 
is proportional to $\delta(y)$, and this makes the $Z$ factor for 
these fields more divergent. From the 4d KK viewpoint, this can be 
seen directly from the one-loop wavefunction diagram. Instead of 
having a sum over a single tower of KK modes in the loop, as in the 
gauge case, there is a double sum, as the KK modes can be different 
in each propagator. Thus in the 4d language, the coefficient of the 
anomalous dimension at energy $E$ is proportional to $(ER)^2$. 
At scale $1/R$, $g_3$ and $y_t$ are comparable, but at larger 
energies the top coupling changes much more rapidly, and is
the first coupling of the theory to become non-perturbative. Paying
careful attention to the thresholds of the different species of
particles, we find that the one-loop evolved top Yukawa coupling
diverges at about $6/R$. Using conventional strong coupling arguments,
the top Yukawa coupling becomes non-perturbative at scale $M$ defined
by
\begin{equation}
  y_t(M) \simeq {4 \pi \over  (MR)^{3/2}}.
\label{eq:M}
\end{equation}
Numerically, $M \simeq 5/R \simeq 1.7~{\rm TeV}$. 
At this scale, the 4d coupling $y_t$ may not be increased relative to 
its value at scale $1/R$ by more than 20\% or so. From the 4d viewpoint 
the loss of perturbativity is largely due to the multiplicity of KK modes.

The gauge couplings change very little over the interval $1/R$ to
$5/R$, and are all perturbative at $M$. At energies larger than $M$, 
evolving at one loop, the gauge couplings become non-perturbative
before unifying --- there is no calculable approximate power law
unification. In fact, the one-loop analysis already becomes unreliable at
$M$, as higher loop diagrams involve the non-perturbative coupling $y_t$.

\subsection{Estimate of brane $Z$ factors}

There are several reasons for studying the size of the brane $Z$
factors:
\begin{itemize} 
\item The compactification scale has UV sensitivity only through the
$Z$ factors. 
\item While the Higgs and top squark masses are remarkably 
insensitive to the $Z$
factors, small changes are possible.
\item As the theory becomes non-perturbative at $M$, is it reasonable
that $z_H(1/R) \lesssim 0.2$?
\item The degeneracy of many states at the
scale $1/R$ could be lifted by even quite small contributions to the
brane $Z$ factors.
\end{itemize}
We can consider two contributions to $Z_H$ -- a boundary value at $M$,
coming from the non-perturbative interactions above $M$, and
a contribution that results from radiative corrections involving $y_t$
from $M$ to $1/R$. 
Here we give rough estimates of these contributions.

Using a strong coupling analysis in higher dimensions \cite{CLP}, 
we estimate an upper bound for the boundary contribution of 
$Z_H \simeq 3 \pi/(2M)$, giving $z_H \simeq 3/(4MR) \simeq 0.2$. 
We do not know that such boundary terms are present;
however, even if this bound is saturated, there is no conflict with 
experiment --- the effects in precision electroweak data are at the
level of current observations. Of course, although not expected, we
cannot exclude large boundary $Z$ factors for $Q_3$ and $U_3$ giving
large shifts to the superpartner masses.

In scaling from $M$ to $1/R$, the
coupling $y_t$ changes by about 20\%. Using the one-loop formulae for
the $Z$ factors, this scaling would generate $z_H \simeq 0.2$. Again
we find a contribution at an acceptable but interesting level. Since
this contribution gets most support from near $M$, 
where the theory approaches strong coupling, it is not
surprising that our two contributions are estimated to be comparable. We
stress that we do not know whether such contributions will be present
in nature. Both are dominated by the UV, and, for example, it could be
that our 5d theory is cut off at some scale below $M$ by being
incorporated into some other theory. Our analysis shows that large
values of $z_H$ are not expected --- our calculation of the
compactification scale is under control and has limited sensitivity 
to the unknown UV physics, and the precise prediction of the Higgs 
mass is maintained.

Light fermions, such as the electron, do not have significant brane
$Z$ factors generated by their Yukawa couplings. However, if our 5d
theory is the low energy effective theory of some more fundamental
theory, valid below some cutoff scale $\Lambda$, then the fundamental
theory could lead to a value of the $Z$ factor at $\Lambda$, given
purely be dimensional analysis, of $|Z| \simeq 1/\Lambda$. In this case
the first superpartner will have a mass shifted to 
$(1/R)( 1 \pm 1/(2 \pi R \Lambda) )$. 
This is a small effect --- a 10 GeV shift for $\Lambda =
M$ --- but represents an important lifting of the degeneracy. 
Brane kinetic terms for the gauge fields could similarly lift the
degeneracy of the gauginos.

\subsection{The $\rho$ parameter}

The $\rho$ parameter has been measured to have the standard model
value, with a limit on additional contributions $\Delta \rho < 0.003$
at 95\% confidence level, providing a powerful probe of new theories.
Unfortunately the $\rho$ parameter cannot be reliably computed in our 
theory.

On first sight it appears that the dominant contribution to $\Delta
\rho$ in our theory comes from the scalar states of the top quark
hypermultiplets which have mass $1/R$ in the absence of EWSB. By
diagonalizing the $2 \times 2$ mass matrix of Eq.~(\ref{mass_1}), we find
that these states contribute $\rho^{(1)} \simeq 0.012$. If this were an
accurate calculation of the dominant contribution to $\Delta \rho$,
our theory would be experimentally excluded. However, it is not a
reliable result for several reasons. If the ``exact'' eigenvalues of
$1/R \pm m_t$ are used we find $\rho^{(1)} \simeq 0.006$. If we
include the ``$3/R$'' scalar states, by diagonalizing the relevant 
$4 \times 4$ scalar mass matrices, we find  $\rho^{(1)} + \rho^{(3)}
\approx 0.04$ using ``exact'' eigenvalues. 
Similarly the fermion states at $2/R$ and $4/R$ are expected to contribute
$\rho^{(2)} + \rho^{(4)} \approx 0.02$. These values are approximate as 
they are sensitive to the precise values used for the masses and mixings.
We conclude that mixing between
the lighter and heavier modes has an important effect on $\rho$. 

The sensitivity of $\rho$ to higher modes is most dramatically seen 
by studying the contribution from the whole KK tower: a quadratic
sensitivity to the UV emerges. We have already seen that radiative
corrections from the top Yukawa interaction  generate the interaction
$\delta(y) \int d^4 \theta Z_H H^\dagger e^{gV} H$ with $Z_H$ having a
quadratic sensitivity to the cutoff due to the power law running of
the top coupling. This $Z$ factor induces a positive contribution,
$\rho_\Sigma$, from the VEV of the electroweak triplet in the 
$\Sigma$ multiplet. 
We now find that the higher dimension operator
$\delta(y) \int d^4 \theta (\xi/ M^4) (H^\dagger e^{gV} H)^2$
is also generated with $\xi$ having quadratic sensitivity to the
cutoff. In the strong coupling limit,
this boundary operator could lead to a contribution as large as
$|\rho^M| \simeq 0.05$. Contributions to the $\rho$ parameter are not 
reliably computed from the lowest lying KK mode.
Our theory requires a cancellation between 
$\rho^{(1)} + \rho^{(2)} + \rho^{(3)} + \rho^{(4)}$ and the 
other contributions to $\Delta \rho$ at the level of one order of 
magnitude.

\section{Superpartner Spectrum and Collider Phenomenology} 
\label{sec:lsp}

What are the first states that will be encountered beyond 
those of the standard model?
From Figure \ref{Fig_spectrum}, we see that these are 
the superpartners, and their $SU(2)_R$ partners, which at tree level 
are all degenerate with mass $1/R$. 
Consider the overall picture of the low energy effective theory
beneath the TeV scale, after the states of mass $2/R$ and above have
been integrated out, so that only the standard model fields and the
``$1/R$'' states remain. It is clear that this effective theory is
very different from the MSSM, which contains 
\begin{itemize}
\item 2 Higgs doublets, and
\item a single set of superpartners.
\end{itemize}
We stress that our effective theory below a TeV contains
\begin{itemize}
\item the 1 Higgs doublet standard model, and
\item two superpartners for every standard model particle.
\end{itemize}
Even though our full theory is supersymmetric, our effective theory is
not. There is no energy scale in which a 4d $N=1$ supersymmetric 
description is appropriate. Rather, the low energy theory reflects 
the underlying $N=2$ supersymmetry coming from the fifth dimension, 
with a superpartner $\tilde{p}$ and a conjugate superpartner 
$\tilde{p}^c$ for every standard model particle $p$
\begin{equation}
  p \Rightarrow (\tilde{p}, \tilde{p}^c).
\label{eq:p}
\end{equation}
In this section, we use a tilde to denote these
superpartners: $\tilde{q}, \tilde{u}, \tilde{d}, \tilde{l}, \tilde{e}$
for the squarks and sleptons, $\tilde{h}$ for the Higgsino, and
$\tilde{g}, \tilde{\omega}, \tilde{z}, \tilde{\gamma}$ for the gauginos,
which we collectively denote by $\tilde{\lambda}$.
The $SU(2)_R$-rotated states\footnote{The only case where $\tilde{p}$
and $\tilde{p}^c$ are not in the same $SU(2)_R$ doublet is the
Higgsino.} we call conjugate superpartners, thus 
$\phi^c_Q = \tilde{q}^c$ is a conjugate squark and $\psi_\Sigma$ 
of the $SU(3)$ gauge multiplet is a conjugate gluino, $\tilde{g}^c$.

\subsection{Spectrum} \label{subsec:sp}

In studying the mass matrices for these states, it is useful to notice
that the theory possesses an accidental, continuous $R$ symmetry. 
We choose it to be such that $\theta$ ($\theta'$) carries $R$ charge
$+1$ ($-1$). The resulting $R$ charges are shown in Table~\ref{R-charge}, 
and are the same for all members of the KK tower. The standard
model particles are neutral, while the superpartners and their
conjugates have charges $\pm 1$.  While the conjugate matter fields
and conjugate gauge fields (i.e. the $\Sigma$ scalar) are neutral, 
the conjugate Higgs field has charge $+2$.
\begin{table}
\begin{center}
\begin{tabular}{|c||c|c|c|} \hline
$R$  & gauge $V$           & Higgs $H$     & matter $M$               \\ \hline
+2   &                     & $h^c$         &                          \\ 
+1   & $\tilde{\lambda} $  & $\tilde{h}^c$ & $\tilde{m}, \tilde{m}^c$ \\ 
0    & $A^\mu, A^c$        & $h$           & $m, m^c$                 \\ 
$-1$ & $\tilde{\lambda}^c$ & $\tilde{h}$   &                          \\ \hline
\end{tabular}
\caption{Continuous $R$ charges for gauge, Higgs and matter components. 
Here, $m$ represents $q, u, d, l, e$.}
\label{R-charge}
\end{center}
\end{table}
The absence of any $A$ terms or Majorana gaugino masses can be traced
to this $R$ symmetry. The pattern of $R$ charges is not symmetrical about 
zero for the matter and Higgs fields because we insist that the Higgs $h$ 
is $R$ neutral, so that the $R$ symmetry is not spontaneously broken.

The leading effect which lifts the degeneracy of these states is EWSB. 
Even though this occurs radiatively, $vR \simeq 0.4$ is given 
by the 4th root of the loop factor $L$, and is not small. 
These effects are important only for the top 
squarks and their conjugates, and for neutralinos and charginos.

There are four top-type scalars, but fortunately $U(1)_R \times
U(1)_{EM}$ ensures the mass matrix
splits into two $2 \times 2$ blocks.
Since $\tilde{t}_L \subset \tilde{q}_3$ and $\tilde{t}_R =
\tilde{u}_3^\dagger$, the two stop scalars $\tilde{t}_L$ and
$\tilde{t}_R$ have opposite $R$ charges and cannot mix. Nevertheless,
mixings can occur within the pair $(\tilde{t}_L,\tilde{t}_R^{c \dagger})$
and the pair $(\tilde{t}_R,\tilde{t}_L^{c \dagger})$. Each pair has
the mass matrix given in Eq.~(\ref{mass_1}).
Taking the mixing with the heavier states into account, the eigenvalues are
\begin{equation}
  m_\pm = {1 \over R} \pm m_t.
\end{equation}
For $R^{-1} = 352$ GeV, there are two charged 2/3 colored scalars, 
$\tilde{t}_{L,R_-}$, of mass 186 GeV, and two, $\tilde{t}_{L,R_+}$, 
of mass 518 GeV. The $L,R$ subscript labels whether the state contains 
$\tilde{t}_L$ or $\tilde{t}_R$.  Each mass eigenstate has
roughly comparable $SU(2)_L$ doublet and singlet components, and each
has a large coupling to the top quark and the Higgsino. Including one-loop 
radiative corrections from the QCD gauge KK tower increases these masses 
by 6\% for $\tilde{t}_-$ to 197 GeV and 1\% for $\tilde{t}_+$ to 522
GeV. Remarkably, the one-loop contribution 
proportional to the top coupling $y_t^2$ vanishes.

\begin{figure}
\centerline{\epsfxsize=11cm \epsfbox{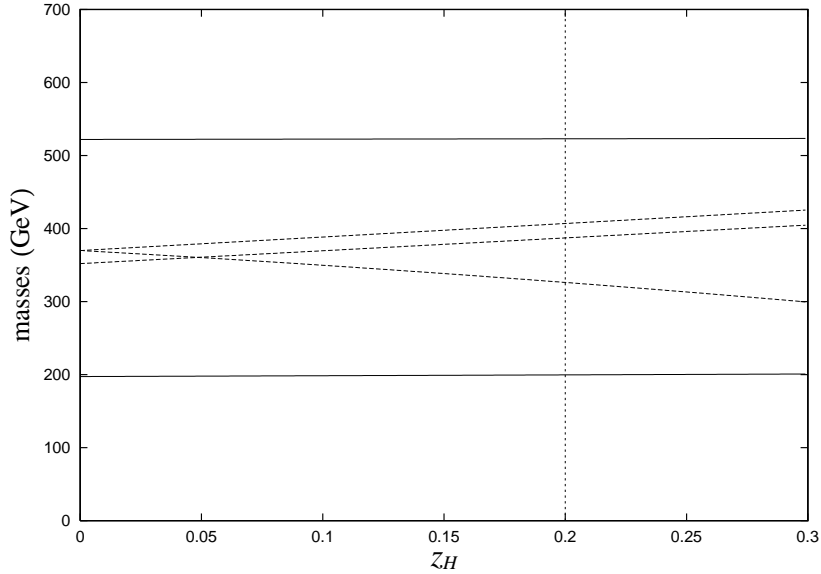}}
\caption{The top-squark and chargino masses as a function of $z_H$.
Solid, dashed lines represent $\tilde{t}_{L,R_{+,-}}$ and 
$\tilde{\chi}^+_{1,2,3}$ fields, respectively.
$z_H > 0.2$ is disfavored by precision electroweak data.}
\label{spectrum-zh}
\end{figure}

Introducing brane $Z$ factors for $Q_3$ and $U_3$, as in 
Eq.~(\ref{eq:braneke}), the $2 \times 2$ mass matrix for the scalars
$(\tilde{t}_L,\tilde{t}_R^{c \dagger})$ becomes
\begin{equation}
  \pmatrix{\tilde{t}_L^\dagger & \tilde{t}_R^c } 
  \pmatrix{ 
    {1-2 z_Q \over R^2} + 4 m_t^2 & 
    - 2 m_t {1-z_U \over R} \cr
    - 2 m_t {1-z_U \over R} & 
    {1-2z_U \over R^2}} 
  \pmatrix{\tilde{t}_L \cr \tilde{t}_R^{c \dagger}}, 
\label{eq:stop2}
\end{equation}
to linear order in $z_{Q,U}$. The mass matrix for the scalars
$(\tilde{t}_R,\tilde{t}_L^{c \dagger})$ is obtained by the interchange 
$z_Q \leftrightarrow z_U$.
To leading order in $z_{Q,U}$ and in $m_t$ the eigenvalues of this 
matrix are $(1-\bar{z})/R \pm m_t$, where $\bar{z} = (z_Q + z_U)/2$. 
Using Eq.~(\ref{eq:boscondz}), which includes mixings with all heavier 
scalar modes, one discovers that this result is exact in $m_t$. 
As shown in section \ref{sec:higgs}, minimization of the Higgs 
potential determines $(1-\bar{z})/R = 352$ GeV, so that the top 
squark masses are independent of the $Z$ factors, to linear order, 
as shown in Figure \ref{spectrum-zh}.

Although there are six charginos and six neutralinos, the mass matrices have
a very simple form, largely determined by 
$U(1)_R$ and $U(1)_{EM}$ symmetries.
Both charginos and neutralinos have a $3 \times 3$ Dirac
mass matrix, which splits into block diagonal form, giving chargino
masses
\begin{equation}
\pmatrix{\tilde{h}^+ & \tilde{\omega}^{c+}} 
\pmatrix{ {1 \over R} &\sqrt{2} M_W \cr
- \sqrt{2} M_W & {1 \over R}} 
\pmatrix{\tilde{h}^{c-} \cr \tilde{\omega}^-} 
+ {1 \over R}\, \tilde{\omega}^{c-} \tilde{\omega}^+,
\label{eq:chargino}
\end{equation}
and neutralino masses
\begin{equation}
\pmatrix{\tilde{h}^0 & \tilde{z}^c} 
\pmatrix{ {1 \over R} & M_Z \cr
-M_Z & {1 \over R}} 
\pmatrix{\tilde{h}^{c0} \cr \tilde{z}} 
+ {1 \over R} \, \tilde{\gamma}^c \tilde{\gamma}.
\label{eq:neutralino}
\end{equation}
The minus signs in the (2,1) entries are extremely important.
The pure gaugino states, $\tilde{\chi}_1^+$ and $\tilde{\chi}_1^0$, 
are the lightest,
with mass $R^{-1} = 352$ GeV. The mixed gaugino/Higgsino states are 
increased in mass: $\tilde{\chi}^+_{2,3}$ both have mass 
$R^{-1} \sqrt{1+2 M_W^2 R^2} = 370$ GeV, 
and $\tilde{\chi}^0_{2,3}$ both have mass 
$R^{-1} \sqrt{1+ M_Z^2 R^2} = 364$ GeV.

These shifts are of order $M_W^2/R$ and hence small. The brane 
wavefunction $Z$ factor for the Higgs field does
not correct the form of the $\tilde{\chi}_1$ masses, but does induce 
corrections to the form of the $\tilde{\chi}_{2,3}$ mass matrices, which, 
to linear order in $z_H$, become 
\begin{equation}
\pmatrix{\tilde{h}^+ & \tilde{\omega}^{c+}} 
\pmatrix{ {1 \over R}(1-z_H) &\sqrt{2} M_W (1+ {1 \over 2} z_H) \cr
- \sqrt{2} M_W  (1- { 1 \over 2} z_H)& {1 \over R}} 
\pmatrix{\tilde{h}^{c-} \cr \tilde{\omega}^-}, 
\label{eq:chargino2}
\end{equation}
for the charginos, and
\begin{equation}
\pmatrix{\tilde{h}^0 & \tilde{z}^c} 
\pmatrix{ {1 \over R}(1-z_H) & M_Z  (1+ {1 \over 2} z_H) \cr
-M_Z  (1- {1 \over 2} z_H)& {1 \over R}} 
\pmatrix{\tilde{h}^{c0} \cr \tilde{z}},
\label{eq:neutralino2}
\end{equation}
for the neutralinos. 
The mass eigenvalues for the charginos are shown in Figure \ref{spectrum-zh} 
as a function of $z_H$. The neutralino mass eigenvalues have a very similar 
behaviour, but with a smaller splitting at $z_H=0$. The $\tilde{\chi}_1^+$ 
state has a mass $R^{-1}$, and changes with $z_H$ only because the
extraction of $R^{-1}$ from the Higgs VEV involves $z_H$. The order
$z_H$ terms in Eq.~(\ref{eq:chargino2}) cause a significant mass splitting
between  $\tilde{\chi}_2^+$ and  $\tilde{\chi}_3^+$, so that
$\tilde{\chi}_1^+$ may well not be the lightest chargino.

All the remaining superpartners, squarks, sleptons, gluinos and their
conjugates, remain closely degenerate, with a mass of $R^{-1}$. 
Radiative corrections from the strong interaction lift the mass of 
the colored states by about 6 GeV compared to the non-colored ones. 
EWSB allows us to determine $R^{-1}= 352 \pm 20$ GeV, where the 
uncertainty is from our theoretical calculation.  Brane $Z$
factors can also affect $R^{-1}$ as shown by the $\tilde{\chi}_1^+$ 
curve in Figure \ref{spectrum-zh},
for the case that the $Z$ factors are generated
purely from the 5d power law running of the top coupling.

The supergravity multiplet in 5d contains the graviton $G_{MN}$, the
gravitino $\Psi_M$ and a vector field $B_M$.
The KK decomposition of this 5d multiplet leads to two 4d
multiplets: the graviton multiplet ($g_{\mu \nu}(+,+), 
\psi_{3/2}(+,-), \psi^c_{3/2}(-,+), B_\mu(-,-)$)
and the radion multiplet ($(g_{55} + i B_5)(+,+), 
\psi_{1/2}(+,-), \psi^c_{1/2}(-,+), g_{\mu 5}(-,-)$).
The towers of KK modes have the usual spectrum with only the 4d graviton
and the radion $g_{55}$ having a zero mode, while the next lightest
states are the fermionic ones at $1/R$. We assume that the radion
acquires a mass by a mechanism that stabilizes the size of the compact
dimension. It has zero $R$ charge and is unstable to decay to standard
model particles.

The two top scalars $\tilde{t}_{L-}$ and $\tilde{t}_{R-}$ are lighter
than all other superpartners by a significant amount, having masses
\begin{equation} 
  m_{\tilde{t}_-} = 197 \pm 20~{\rm GeV}.
\label{eq:lsp}
\end{equation}
This result is remarkably insensitive to brane $Z$ factors, as shown
in Figure \ref{spectrum-zh}. The top scalars
remain at least 100 GeV lighter than the lightest chargino or
neutralino. In addition, the brane $Z$ factors do not lift the 
degeneracy between $\tilde{t}_{L-}$ and $\tilde{t}_{R-}$ 
to the leading order.  Since the one-loop gauge correction and $D$-term 
contribution also do not lift it, the mass splitting between these 
two scalars would be, if any, very small.

Although $\tilde{t}_{L-}$ and $\tilde{t}_{R-}$ have opposite 
$U(1)_R$ charges, the heavier may decay to the lighter, for example by
$\tilde{t}_{L-} \rightarrow \tilde{t}_{R-}^\dagger uu$. This decay could be
mediated by a flavor-changing gluino exchange, and the lifetime is
very sensitive to the scalar mass differences and the size of the
flavor changing coupling. Decays may also be induced by certain higher
dimension operators. For the most plausible ranges of parameters the
decay does not occur inside a collider detector, unless the
heavier scalar is stopped, but would be expected to occur
cosmologically. We refer to these two scalars as the LSP and 
the next to LSP (NLSP).

\subsection{Collider phenomenology}

For run II of the Fermilab collider the most promising signals 
of our theory are the standard model Higgs of mass at $127 \pm 8$
GeV, and the LSP and NLSP top scalars of mass $197 \pm 20$ GeV. The 
pair production cross section for these scalars at a center of mass energy
of 2 TeV is $360~{\rm fb}$ each \cite{BKPS}. 
This is significantly more than the rate 
for producing them via pair production of the gluinos and other squarks
with mass of 350 GeV, followed by cascade decays to the LSPs. Once
pair produced, these scalars will hadronize by picking up a $u$ or $d$
quark and becoming a fermionic meson $T^0 = \bar{u} \tilde{t}_-$ or
$T^+ = \bar{d} \tilde{t}_-$, with almost equal probability. 
While the charged meson is expected to be slightly heavier than the 
neutral one, both will be sufficiently stable to traverse the entire 
detector. Hence the signals for scalar top pair production are events 
with one or two heavy stable particles with electric charge 
$\pm e$.\footnote{
If $R$-parity violating interactions are introduced on the branes, 
the top scalars may decay into the standard model particles inside 
the detector.}
Furthermore, the anti-stop bound states, 
$\bar{T}^0 = u \bar{\tilde{t}}_-$ and $T^- = d \bar{\tilde{t}}_-$, 
can oscillate by exchanging isospin and charge with background 
material in the detector, causing intermittent highly ionizing tracks.  
These signals have been investigated in the context of gauge 
mediation models with stop NLSP \cite{stopNLSP}. 
The present experimental limit on such particles from Run Ib at the 
Tevatron collider is about 150 GeV from CDF \cite{CDF}. 

For sufficiently low speed $\beta$, the
charged mesons $T^+$ will stop inside the detector, and eventually
give a positron with very low momentum, presumably in the MeV range, 
by $\beta$ decay to the neutral state. 
If $T^+$ contains the NLSP, another possibility is that 
this NLSP decays to the LSP giving decay products with energies 
of order the scalar mass difference, for example in the GeV range.

At the LHC the situation is more complicated since all the squarks,
gluinos and their conjugates will be pair produced. However, these
will all cascade to the LSP or NLSP, so once again a crucial signal becomes
the observation of events with one or two stable charged particles.
The initial pair production reaction produces one particle with $U(1)_R$
charge $+1$, and one with $-1$. These cascade to eventually give  
$\tilde{t}_{L-}$ or $\tilde{t}_{R-}^\dagger$ and
$\tilde{t}_{R-}$ or $\tilde{t}_{L-}^\dagger$ respectively. Thus,
unlike in the case of direct scalar top production, events
occur in which the stable charged particles have the same sign.

If the gluino is heavier than the squarks the decay chain is 
$\tilde{g} \rightarrow q \tilde{q}$ and $\tilde{q} \rightarrow 
q \tilde{\chi}$, where $q$ and $\tilde{q}$ refer to any flavor
except top, when the decay is not forbidden by phase space. For the 
case that the squarks are heavier than the gluino the decays are 
$\tilde{q} \rightarrow q \tilde{g}$  and $\tilde{g} \rightarrow 
q \bar{q} \tilde{\chi}$. Recall that $\tilde{g}$ and $\tilde{\chi}$ 
are Dirac fermions, and $\tilde{q}$ refers to squarks and 
conjugate squarks. All $\tilde{\chi}$ states that are kinematically 
open will be populated. 
The three charginos will all decay via $\tilde{\chi}^+ \rightarrow 
\bar{b} \tilde{t}_-$. The neutralinos decay to $\bar{t} \tilde{t}_-$ 
if open, but the lightest neutralino will decay via a virtual chargino: 
$\tilde{\chi}^0 \rightarrow \bar{q} q \tilde{\chi}^+$,
$\tilde{\chi}^+ \rightarrow \bar{b} \tilde{t}_-$. Hence all
events resulting from pair production of the ``$1/R$'' states 
contain two $b$ quark jets.

\subsection{U(1)$_R$ symmetry}

In subsection \ref{subsec:sp}, an accidental $U(1)_R$ symmetry, 
given in Table~\ref{R-charge}, played an important role to reveal 
the mass spectrum of the theory.
Here we note that this $U(1)_R$ is an anomaly-free symmetry.
In the usual 4d $N=1$ theory, an anomaly free $U(1)_R$ symmetry 
requires the introduction of additional exotic states, 
beyond those of the MSSM, since the gauginos and the gravitino carry a 
$U(1)_R$ anomaly \cite{CDCFM}.
However, in the present model, there are conjugate gauginos and
conjugate gravitino, so that the $U(1)_R$ is automatically
anomaly free; the quarks and leptons have zero charges and 
all the other fermionic states are vector-like.
The $U(1)_R$ symmetry is a linear combination 
of the $U(1)$ subgroup of the $SU(2)_R$ automorphism group of $N=2$ 
supersymmetry algebra and a vector-like, non-$R$ $U(1)$ symmetry 
under which the Higgs fields transform as $H(-1)$ and $H^c(+1)$ 
($H'(-1)$ and $H^{c\,\prime}(+1)$) again demonstrating that 
$U(1)_R$ is anomaly free.

The above remarkable property allows us to impose the $R$ symmetry 
as an exact symmetry of the theory.
Since neither EWSB $\left\langle h \right\rangle \neq 0$ 
nor chiral condensation $\left\langle q \bar{q} \right\rangle \neq 0$ 
breaks this $U(1)_R$ symmetry, it may remain as an unbroken symmetry.
Then, the LSP is absolutely stable since the $R$ parity, $R_P$ 
or $R_P'$, is a discrete subgroup of the $U(1)_R$ symmetry, and 
some higher-dimensional operators are forbidden by the symmetry.

Anomaly freedom raises the interesting
possibility that this $U(1)_R$ symmetry is a gauge symmetry.
In the usual 4d $N=1$ supergravity, a gauged $R$ symmetry cannot remain
at low energies since a non-vanishing Fayet-Iliopoulos $D$ term of order
the Planck scale is generated and breaks $U(1)_R$ at the Planck 
scale \cite{DZF}.
This is true even in the $N=2$ case \cite{BS}.
If the same is true of our theory, the low-energy consequence of 
the gauged $U(1)_R$ may only be the presence of a discrete gauge 
symmetry such as $R$ parity.
However, it is not completely clear whether the $U(1)_R$ symmetry is 
necessarily broken in the present theory, and if it remains unbroken 
at low energies we would have a massless $U(1)_R$ gauge boson which 
couples only to the superpartners and conjugate superpartners.

\section{Conclusions} \label{sec:concl}

In this paper we have proposed an embedding of the standard model in a
supersymmetric theory with an extra dimension compactified on the
orbifold $S^1/(Z_2 \times Z_2')$ --- a circle with two orthogonal
reflection symmetries. All standard model particles propagate in the
bulk, yet, because the compactification breaks supersymmetry,
they are the only massless modes at tree level.  The only interactions 
in the bulk are the 5d supersymmetric $SU(3) \times SU(2) \times U(1)$ 
gauge interactions, which lead at low energies to the
standard model gauge interactions and the tree-level Higgs potential 
$(g^2 + g'^2) |\phi_H|^4 / 8$.
The orbifold allows for two different types of hypermultiplet, according 
to whether the zero mode is a scalar or a fermion.  Thus the orbifold 
provides a distinction between the Higgs and lepton doublet superfields. 
 
The Yukawa couplings for up-type quarks appear as 4d supersymmetric
interactions  at the fixed points of the $Z_2$ symmetry, while
the Yukawa couplings for down-type quarks and charged leptons appear
as 4d supersymmetric interactions  at the fixed points of the $Z_2'$
symmetry. Because supersymmetry acts differently at these spatially
separated locations, all Yukawa interactions involve just a single Higgs
doublet,  $\phi_H$. The top quark KK tower induces a one-loop effective
potential for the Higgs which depends on the compactification scale.
This potential contains a negative mass-squared, and is a
monotonically decreasing function of $|\phi_H|$. It triggers EWSB, but
runaway behaviour is prevented by the stabilizing effect of the
tree-level gauge quartic interaction. Requiring the minimum to give the
observed value for the Fermi coupling determines
the compactification scale: $R^{-1} = 352 \pm 20$ GeV, 
leading to a prediction for the Higgs mass of $m_H = 127 \pm 8$ GeV.
It is remarkable that the Higgs potential does not have the standard 
model form, and the dependence of the top quark mass on the Higgs VEV 
is also non-standard. 

The nature of the orbifold automatically ensures that supersymmetry is
broken and that the superpartners all have mass $R^{-1}$ at tree level.
There is no additional supersymmetry breaking sector or mediation
mechanism, and hence no soft supersymmetry breaking parameters such as
$\mu, m_{1/2}, m_0, A$ or $B$. There is no fine tuning between parameters 
for successful EWSB, as there is only a single free parameter, $R^{-1}$.
Because supersymmetry is broken non-locally by the global properties of
the orbifold, all supersymmetry breaking effects are finite and reliably
calculated in terms of physics at the compactification scale. Unlike 4d
supersymmetric theories, which have a large logarithmic dependence on
physics
at high energies, supersymmetry breaking effects are exponentially
shielded from physics above $R^{-1}$.
 
There are three Dirac charginos of mass (352, 370, 370) GeV and three
Dirac neutralinos of mass (352, 364, 364) GeV, where the splitting is
induced by EWSB. The EWSB splittings are largest for the top squarks,
with two having a mass of $197 \pm 20$ GeV and two having a mass of $522$
GeV. All other superpartners have a mass close to $R^{-1}$, while all KK
excitations of the standard model have a mass close to  $2 R^{-1} = 
704 \pm 40$. The LSP is a top squark and is most likely stable.  It 
will be copiously produced at future runs of the Tevatron collider.

Sensitivity to physics at energies much above $R^{-1}$ enters only
through supersymmetric interactions at the orbifold fixed points. This
gives an additional 20\% uncertainty to all the masses given above, 
except for the masses of the top squarks and the Higgs boson, 
which are more rigidly tied to EWSB. 
Higher dimension operators also contribute to the $\rho$ parameter, 
and must cancel contributions from the superpartners and KK excitations 
at the level of one order of magnitude. This UV sensitivity implies 
that $\rho$ cannot be reliably computed.

Our theory becomes non-perturbative at a scale of about 2 TeV, 
above which it becomes incorporated in some other higher dimensional theory. 
This might be a quasi-conformal field theory with an energy desert, 
or it could be that the fundamental scale of gravity is not far above 
2 TeV \cite{ADD}.

\section*{Acknowledgements}

We would like to thank N.~Arkani-Hamed, R.~Rattazzi, D.~Smith 
and N.~Weiner for discussions.
Y.N. thanks the Miller Institute for Basic Research in Science 
for financial support.
This work was supported by the E.C. under the RTN contract 
HPRn-CT-2000-00148, the Department of Energy under contract 
DE-AC03-76SF00098 and the National Science Foundation under 
contract PHY-95-14797.

\newpage

\def\pl#1#2#3{{\it Phys. Lett. }{\bf B#1~}(19#2)~#3}
\def\zp#1#2#3{{\it Z. Phys. }{\bf C#1~}(19#2)~#3}
\def\prl#1#2#3{{\it Phys. Rev. Lett. }{\bf #1~}(19#2)~#3}
\def\rmp#1#2#3{{\it Rev. Mod. Phys. }{\bf #1~}(19#2)~#3}
\def\prep#1#2#3{{\it Phys. Rep. }{\bf #1~}(19#2)~#3}
\def\pr#1#2#3{{\it Phys. Rev. }{\bf D#1~}(19#2)~#3}
\def\np#1#2#3{{\it Nucl. Phys. }{\bf B#1~}(19#2)~#3}

\end{document}